\pdfoutput=1
\documentclass[10pt,letterpaper]{article}

\usepackage[T1]{fontenc}
\usepackage[utf8]{inputenc}
\usepackage{newtxtext,newtxmath}
\usepackage{microtype}
\usepackage[letterpaper,top=0.72in,bottom=0.72in,left=1.0in,right=1.0in,
            headheight=14pt,headsep=12pt,footskip=24pt]{geometry}
\usepackage{amsmath,mathtools}
\usepackage{graphicx}
\usepackage{booktabs,longtable,tabularx,array,multirow}
\usepackage{enumitem}
\usepackage{caption}
\usepackage{pdflscape}
\usepackage{float}
\usepackage{needspace}
\usepackage{xcolor}
\usepackage{xurl}
\usepackage[backend=biber,style=ieee,sorting=none,doi=true,url=true,isbn=true,maxbibnames=999]{biblatex}
\AtBeginBibliography{\small}
\usepackage{titlesec}
\usepackage{fancyhdr}
\usepackage[hidelinks]{hyperref}
\usepackage{bookmark}

\renewcommand{\arraystretch}{1.12}
\newcolumntype{L}[1]{>{\raggedright\arraybackslash}p{#1}}
\newcommand{\IndexedAnd}[1]{\mathop{\scalebox{1.20}{$\wedge$}}\limits_{#1}}
\setlist[itemize]{leftmargin=1.8em,itemsep=0.1em,topsep=0.2em,parsep=0pt}
\setlist[enumerate]{leftmargin=2.0em,itemsep=0.1em,topsep=0.2em,parsep=0pt}
\titleformat{\section}{\normalfont\bfseries\large}{\thesection}{0.75em}{}
\titleformat{\subsection}{\normalfont\bfseries\normalsize}{\thesubsection}{0.65em}{}
\titleformat{\subsubsection}{\normalfont\bfseries\normalsize}{\thesubsubsection}{0.65em}{}
\titleformat{\paragraph}[runin]{\normalfont\bfseries}{ }{0pt}{}
\titlespacing*{\section}{0pt}{1.2em}{0.45em}
\titlespacing*{\subsection}{0pt}{0.9em}{0.35em}
\titlespacing*{\subsubsection}{0pt}{0.75em}{0.3em}
\titlespacing*{\paragraph}{0pt}{0.65em}{0.55em}

\fancypagestyle{preprint}{%
  \fancyhf{}
  \fancyhead[L]{\small Software Engineering in the Agent Era}
  \fancyhead[R]{\small\scshape A Preprint}
  \fancyfoot[C]{\small\thepage}

}
\hypersetup{
  pdftitle={Software Engineering in the Agent Era: From Trustworthy Change to Human--Agent Software Organizations},
  pdfauthor={Zhongjie Wang; Mingyi Liu},
  pdfsubject={Agentic Software Engineering},
  pdfkeywords={Agent Software Engineering, Trustworthy Change, Human--Agent Cell, Responsibility Topology}
}

\begin{document}
\thispagestyle{empty}

\begin{center}
\rule{\textwidth}{1.0pt}
\vspace{0.75em}

{\Large\bfseries\scshape
Software Engineering in the Agent Era:\\[0.15em]
From Trustworthy Change to Human--Agent Software Organizations\par}

\vspace{0.7em}
\rule{\textwidth}{1.0pt}
\vspace{0.8em}

{\small\scshape A Preprint\par}
\vspace{1.15em}

\begin{tabular*}{\textwidth}{@{\extracolsep{\fill}}cc@{}}
\begin{minipage}[t]{0.44\textwidth}\centering
\textbf{Zhongjie Wang}\\[-0.05em]
Faculty of Computing\\
Harbin Institute of Technology\\
China\\
\texttt{rainy@hit.edu.cn}
\end{minipage}
&
\begin{minipage}[t]{0.44\textwidth}\centering
\textbf{Mingyi Liu}\\[-0.05em]
Faculty of Computing\\
Harbin Institute of Technology\\
China\\
\texttt{liumy@hit.edu.cn}
\end{minipage}
\end{tabular*}

\vspace{1.2em}
September 4, 2026
\end{center}

\vspace{0.65em}
\begin{center}\bfseries ABSTRACT\end{center}
\vspace{-0.55em}
\small
Software agents are making one part of software engineering unusually elastic: digital execution. Repository analysis, code generation, testing, migration, tool use, and portions of operations can increasingly be replicated and parallelized without proportional growth in human headcount. Other constraints do not scale in the same way. Problem framing, semantic commitment, verification, integration, attention, and acceptance of residual risk remain bounded by human cognition, organizational authority, and economic capacity. The resulting asymmetry raises a software-engineering question that is broader than code generation: how should scalable execution be governed so that it produces software changes that an organization can accept and sustain through their lifecycles?

We develop a testable framework around two primary theoretical constructs and one execution abstraction. \textbf{Trustworthy Change (TC)} is the engineering object that moves from intent through delegated execution, verification, integration, acceptance, and operation. \textbf{Responsibility Topology} classifies a software organization by the distribution of independent authority to accept residual risk. A \textbf{single-center} topology has one final responsibility anchor for the baseline; a \textbf{multi-anchor} topology requires joint acceptance across independently governed domains. The \textbf{Human--Agent Cell (HAC)} is an execution abstraction: it produces candidates, proposals, and evidence, but execution alone does not confer acceptance authority. This separation becomes consequential as execution and acceptance authority scale along different dimensions: distributed HAC execution creates context-coherence and invalidation pressures, while multi-anchor governance adds joint acceptance and explicit responsibility-closure requirements.

Responsibility, accountability, change management, specification, verification, and human oversight all predate this paper. Our claim is narrower: agent-scaled execution changes how these concerns fit together. We make independent residual-risk acceptance authority an explicit organizational classification axis and derive consequences for change state, shared engineering facts, verification, and flow control. Progressive Specification and bounded-capacity analysis remain hypotheses to be tested rather than laws asserted by the framework. The contribution is theory construction and operationalization; empirical validity remains open to controlled, longitudinal, and field studies.

\vspace{0.45em}
\noindent\textbf{\textit{Keywords}}\quad Agent Software Engineering $\cdot$ Trustworthy Change $\cdot$ Human--Agent Cell $\cdot$ Responsibility Topology $\cdot$ Single-Center Software Engineering $\cdot$ Multi-Anchor Team Software Engineering $\cdot$ Progressive Specification $\cdot$ Human--Agent Software Organization
\normalsize

\section{Introduction: The Scale of Software Execution Is Changing}
\label{sec:introduction}

\subsection{Scalable Digital Execution and Shifting Scarcity}
Many mainstream software-development planning and organizational mechanisms have historically treated human execution capacity as a major constraint.

Compilers, IDEs, automated testing, continuous integration, and cloud infrastructure steadily raised developer productivity, but requirements analysis, program construction, fault diagnosis, verification, and maintenance continued to consume substantial human time. Consequently, whether a project followed waterfall development, agile methods, DevOps, or large-scale engineering practices, planning, task decomposition, organization design, and cost estimation were ultimately constrained by a scarce resource:
\[
\text{Human Effort}.
\]

Software agents weaken this assumption. A single developer can now delegate multiple digital execution processes to analyze a repository, generate code, construct tests, locate defects, migrate APIs, process data, configure environments, update documentation, and perform parts of deployment and runtime diagnosis. Relative to conventional human labor, this execution capacity has three distinctive properties.

The distinctive change is not simply that an agent writes code faster. Digital execution can be replicated, run concurrently in isolated workspaces, and expanded through token, compute, and tool budgets on a timescale far shorter than recruiting and training a human team. Its supply is now more elastic than the organizational mechanisms that decide what should be built and whether the result should be accepted.

The formerly stable relationship
\[
\text{Execution Capacity}\propto\text{Number of Human Developers}
\]
is weakening.

SE~3.0 frames this transition as an intent-oriented, conversational form of human--AI software engineering~\cite{ref8}; SASE further places agents within the Actor, Process, Tool, and Artifact structure of software engineering~\cite{ref9}. By 2026, mainstream agent-development environments already support long-running tasks, concurrent agents, isolated workspaces, and programmable tool use. The first-order effect is a restructuring of the \emph{supply mechanism for software execution}.

\paragraph{Shifting scarcity.}
The other constraints of software development remain. In the sense of Brooks's classic distinction between essential and accidental difficulties, agents may compress substantial implementation work without eliminating the semantic and organizational difficulty of deciding what a system should do and what evidence is sufficient to accept it~\cite{ref1}. An agent may implement a change rapidly, yet code completion leaves open whether the development goal is sufficiently decided, whether material ambiguity remains, whether a local implementation satisfies system-level constraints, whether verification evidence justifies the required claims, whether multiple local changes compose safely, and who can accept the residual risk after release.

We use one running example throughout the paper. An order system initially defines
\[
\texttt{status}=\{\texttt{active},\texttt{cancelled}\},
\]
where \texttt{cancelled} conflates two meanings: user-initiated cancellation and automatic closure after payment timeout. The system later needs
\[
\texttt{status}=\{\texttt{active},\texttt{cancelled},\texttt{expired}\},
\]
with
\[
\texttt{cancelled}=\text{user-initiated cancellation},
\]
\[
\texttt{expired}=\text{automatic closure caused by payment timeout}.
\]

At the code level, the change may look like one new enumeration value and a few conditional branches. Agents can propagate it quickly through domain models, payment services, SDKs, tests, analytics, and documentation. The engineering problem is larger. What happens if a successful payment callback arrives after timeout? Does the new status interact with refunds already in progress? Can historical data distinguish the two closure reasons without unsupported inference? How should an old SDK behave when it encounters the new enum? When does analytics adopt the new semantics? Which responsible roles can accept the altered order and payment behavior?

The agent era increasingly exhibits
\[
\text{Execution Capacity}\gg\text{Judgment, Verification, Integration, and Responsibility}.
\]
As implementation capacity grows, bottlenecks move toward semantic decisions, verification, integration, resource management, and formal acceptance of risk.

\subsection{From Human Process Control to Trusted Change Governance}
A substantial part of software-engineering process design has historically organized and constrained scarce human execution capacity. Requirements, design, task decomposition, review, testing, configuration management, and project management constrain human activity and turn many local actions into maintainable software systems.

\begin{figure}[H]
\centering
\includegraphics[width=0.98\linewidth]{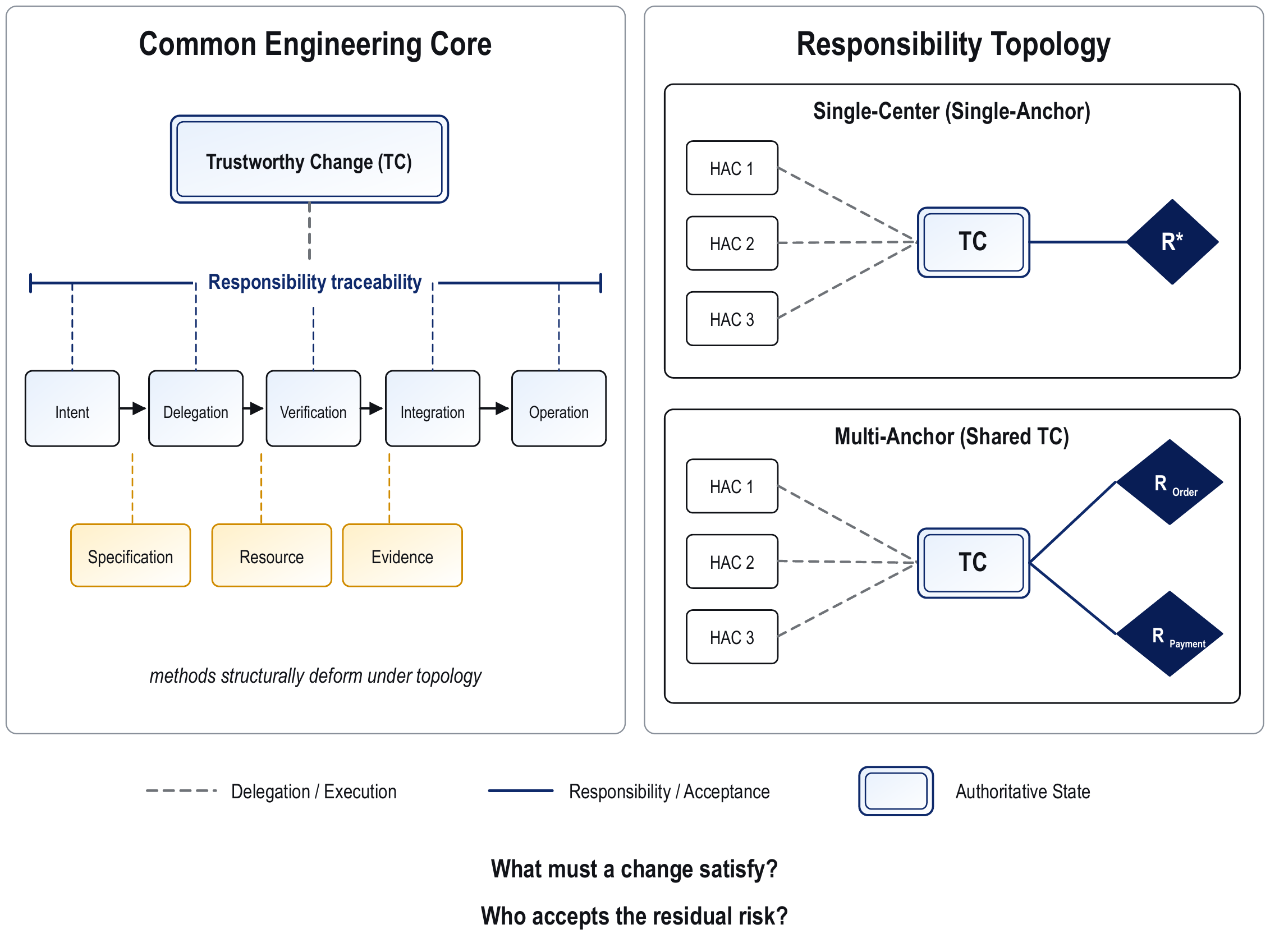}
\caption{The two basic questions of Agent Software Engineering: what engineering conditions a software change must satisfy, and who can accept its residual risk.}
\label{fig:overall-framework}
\end{figure}

These mechanisms remain necessary in agentic development, but the object of control must expand. One software change may be carried out by multiple agents, tasks, and commits. Merely observing people, agent tasks, or code submissions no longer tells us whether that change has satisfied the conditions required to enter an authoritative baseline. We summarize the shift as
\[
\text{Human Process Control}\rightarrow\text{Trusted Change Governance}.
\]
We use \textbf{Trusted Change Governance} for the governance paradigm and \textbf{Trustworthy Change} for the software change whose engineering acceptability is governed. Trusted Change Governance follows a change from intent formation and delegated execution through verification, integration, acceptance, and runtime operation. Figure~\ref{fig:overall-framework} summarizes these two questions and previews the responsibility structures formalized in Section~\ref{sec:topology}: a single-center form with one final anchor $R^*$, and a multi-anchor form illustrated by the Order and Payment anchors.

The resulting theoretical grammar is deliberately compact:
\[
\boxed{\text{Agent SE}=\text{Common Engineering Core}+\text{Responsibility Topology}}.
\]
The common engineering core describes what a software change must satisfy; Responsibility Topology describes the authority structure through which those conditions are finally accepted.

\subsection{Research Positioning, Questions, and Contributions}
Current agentic software engineering spans several overlapping directions. SE~3.0 and SASE examine changes to actors, processes, tools, and artifacts when agents enter the lifecycle~\cite{ref8,ref9}. MAGE argues that as implementation becomes more abundant, engineering scarcity shifts toward representation, evidence, and authority~\cite{ref10}. Hoda explicitly argues that Agentic SE must be framed beyond code and across the whole software process~\cite{ref44}; Alenezi similarly emphasizes verification bottlenecks, accountable oversight, and a shift from code authorship toward outcome ownership~\cite{ref42,ref43}. Against this background, our contribution is narrower. Whole-process scope, accountability, and human oversight are established concerns; we focus on how independent residual-risk acceptance authority structures software organizations and change governance.

Spec-Driven Development and harness engineering, meanwhile, move agentic development away from one-shot prompts toward persistent specifications, rules, skills, hooks, MCP-based tools, sandboxes, and agent runtimes. Microsoft advocates structured specifications as a shared engineering basis for humans and AI, while GitHub Spec Kit makes Spec, Plan, Task, and Implementation explicit elements of the development process~\cite{ref21,ref22}.

Trust-oriented work is also shifting evaluation from whether an agent completes a task to whether success is sufficiently justified. Programming with Trust, Trustworthy AI Software Engineers, and process-oriented agent evaluation investigate evidence, inspection, accountability, and process quality~\cite{ref11,ref12,ref14}.

A second lineage predates current coding agents and comes from software-engineering questions about organization, competing interests, and controlled evolution. Earlier work by the authors examined role distribution and transformation in software project teams~\cite{ref51}, multi-stakeholder value/quality conflicts in composite services~\cite{ref52}, and runtime adaptation and evolution in microservice systems~\cite{ref55}. Those settings differ substantially from agentic software development, but they repeatedly expose the same structural tension: locally acting units must retain enough autonomy to do useful work while the system as a whole still needs a defensible basis for coordination and acceptance. Agentic execution changes the speed and scale of that tension. Responsibility Topology addresses the authority dimension, while Authoritative Engineering State and Context Invalidation address coordination of distributed execution around binding engineering facts.

Responsibility Topology is orthogonal to the SE4H/SE4A distinction. SE4H/SE4A asks whether software engineering activity is primarily oriented toward humans or agents; Responsibility Topology asks how residual-risk acceptance authority is organized. A single-center organization can have both SE4H and SE4A capabilities, as can a multi-anchor team. Actor or engineering-mode classifications cannot substitute for the authority topology.

\paragraph{Relation to established governance mechanisms.}
Traditional mechanisms such as RACI-style role matrices, configuration/change approval boards, CODEOWNERS, branch or release approval workflows, and separation-of-duty rules are adjacent antecedents rather than substitutes for Responsibility Topology. A role matrix may record who is responsible, consulted, or accountable without specifying whether one actor can unilaterally override another's final residual-risk acceptance right. Multiple CODEOWNERS or approval steps can still sit under one final responsibility center; conversely, a safety or compliance authority whose required acceptance cannot normally be waived constitutes an independent anchor for affected changes. Responsibility Topology therefore classifies effective acceptance authority rather than workflow participation, repository permission, or approval count.

Our central question is consequently how \emph{execution, software change, shared engineering facts, and formal acceptance authority} should be represented in one software-engineering model once agents become first-class execution actors.

\paragraph{Research questions.}
The paper addresses four research questions.

\textbf{RQ1.} Which production constraints of traditional software engineering are changed by agents, and why does this require engineering control to shift from human activity toward the software change itself?

\textbf{RQ2.} How does Responsibility Topology distinguish single-center SE from multi-anchor Team SE, and why can the distinction not be reduced to human or agent headcount?

\textbf{RQ3.} Which methods form a common engineering core for Agent SE, and which coordination and closure mechanisms arise from distributed execution versus Responsibility Topology?

\textbf{RQ4.} Under bounded token, compute, tool, wall-clock, human-attention, and verification capacity, how should the production of Trustworthy Change be measured, estimated, and managed?

\paragraph{Contributions.}
Table~\ref{tab:contributions} summarizes the main contributions and their theoretical status.
\begin{table}[H]
\centering
\small
\caption{Main contributions of the paper.}
\label{tab:contributions}
\begin{tabularx}{\linewidth}{@{}p{0.18\linewidth}p{0.32\linewidth}X@{}}
\toprule
\textbf{Type} & \textbf{Contribution} & \textbf{Core question} \\
\midrule
Classification construct & \textbf{Responsibility Topology} & How is authority to accept residual software risk distributed? \\
Execution abstraction & \textbf{Human--Agent Cell (HAC)} & What is the execution-participation unit in agentic software work? \\
Integrative construct & \textbf{Trustworthy Change (TC)} & What constitutes a change that may formally enter the software baseline? \\
Derived mechanisms & \textbf{Authoritative State / Context Invalidation / Responsibility Closure} & How do distributed execution and distributed authority create different coordination and closure requirements? \\
Method hypothesis & \textbf{Progressive Specification Engineering} & How far should software intent be refined before delegated execution? \\
Research program & \textbf{Bounded-Capacity Agent SE} & How do bounded digital resources and verification capacity constrain trustworthy throughput? \\
\bottomrule
\end{tabularx}
\end{table}

The contributions do not all have the same theoretical status. Responsibility Topology introduces a new organizational classification axis. Authoritative State, Context Invalidation, and Responsibility Closure are derived coordination and closure mechanisms: distributed execution creates context-divergence and coordination pressures, while independent anchors determine formal closure conditions. HAC and TC are integrative constructs for agentic execution: they reorganize established concerns in human--AI collaboration, change management, evidence, resources, and responsibility into constructs with explicit boundaries and testable consequences. Progressive Specification and Bounded-Capacity Agent SE play the roles of a method hypothesis and a measurement/economics research program, respectively. Tools, metrics, agent hierarchies, and claims--evidence graphs are supporting mechanisms rather than parallel theoretical layers.

The novelty claim is intentionally narrower than the vocabulary may suggest. Responsibility, accountability, human oversight, software change, specification, and verification all have long histories in software engineering and governance. Those concerns are prior work. The organizational contribution is the use of the \emph{distribution of independent residual-risk acceptance authority} as a primary classification axis for Agent SE; the object-level contribution is the explicit Candidate--Eligible--Accepted change lifecycle with resource, evidence, runtime, and responsibility closure attached to the same change. This narrower claim is important because several contemporary approaches already address human--agent responsibility boundaries, approval protocols, graduated oversight, and engineering risk without using the same classification axis~\cite{ref41,ref49,ref50}.

\subsection{Theory Status, Scope, and Empirical Commitments}
We distinguish three stages of theory development:
\[
\text{Conceptualization}\rightarrow\text{Operationalization}\rightarrow\text{Empirical Testing}.
\]
The present framework concentrates on the first two: it defines constructs and their relations, and it specifies how they can be mapped to observable engineering phenomena. Section~\ref{sec:agenda} states empirical commitments and falsification conditions. Internal coherence is a quality criterion for theory formulation, not a substitute for empirical validity~\cite{ref33,ref34}.

\begin{figure}[H]
\centering
\includegraphics[width=0.94\linewidth]{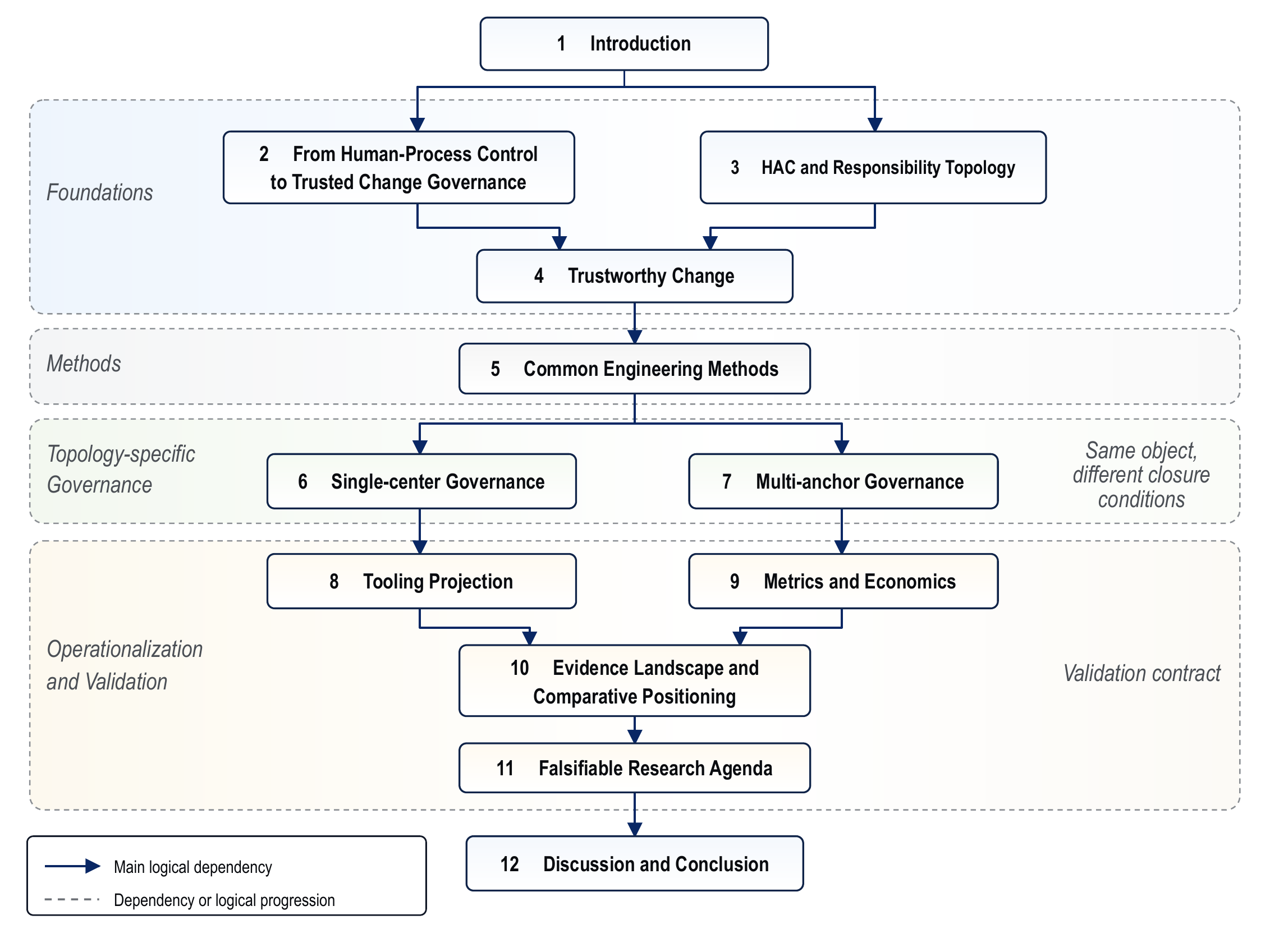}
\caption{Roadmap of the paper and logical dependencies among its major parts.}
\label{fig:paper-roadmap}
\end{figure}

\paragraph{Paper organization.}
Figure~\ref{fig:paper-roadmap} summarizes the logical structure of the paper. Section~2 analyzes the shift in the control object of software engineering. Section~3 defines HACs, Responsibility Anchors, and Responsibility Topology. Section~4 formalizes Trustworthy Change and acceptance. Section~5 presents the common engineering methods. Sections~6 and~7 study the single-center and multi-anchor forms. Section~8 gives a reference tooling architecture, Section~9 addresses measurement and economics, Section~10 maps the research and industrial landscape as of September 2, 2026, Section~11 defines empirical commitments and a falsifiable research agenda, and Section~12 discusses longer-term implications.

\section{From Human Process Control to Trusted Change Governance}
\label{sec:governance}

\subsection{Production Constraints and Capability Mismatch}
Software engineering has always extended beyond programming. Historically, it has served three broad functions.

First, it constrains development activity. Requirements, design rules, coding standards, reviews, testing, configuration management, and change control reduce the freedom of local action so that complex systems can remain coherent across many developers and long periods of evolution.

Second, it augments development capability. Compilers, IDEs, build tools, automated tests, CI/CD, and cloud platforms continuously reduce the mechanical cost of software construction and operation.

Third, it organizes collaboration. Roles, plans, module boundaries, ownership, configuration baselines, meetings, and project metrics turn many local contributions into system-level results.

The Personal Software Process and Team Software Process are systematic examples of process discipline at individual and team scales~\cite{ref4,ref5}. From this perspective, the persistent problem of traditional software engineering can be summarized as
\[
\text{Limited Human Capacity}\rightarrow\text{Reliable Software}.
\]

\paragraph{Agentic execution.}
Agents reduce the marginal cost of search, generation, modification, testing, and local diagnosis. Execution capacity that previously required additional staffing can increasingly be expanded through model calls, agent instances, and compute resources.

The change primarily affects the execution side of software work. The causal constraints of software systems do not disappear. State, authorization, data consistency, interface compatibility, performance, and security still depend on system behavior, not on whether the implementation was produced by a human or an agent.

MAGE characterizes this transition as a relative scarcity of engineering judgment once implementation capacity expands, and highlights representation, evidence, and acceptance obligations~\cite{ref10}. A 12-week case study from the same research line develops a candidate middle-range theory of governance conversion: recurring failures exposed by high-velocity agentic implementation are converted into durable engineering controls~\cite{ref59}. Independent work on Agentic SE likewise identifies verification as a critical quality bottleneck once code generation becomes abundant~\cite{ref43}. The distribution of value in software engineering therefore shifts: implementation remains necessary, but intent, specification, verification, integration, and responsibility occupy a larger share of the complete change cost.

\paragraph{Capability mismatch.}
Table~\ref{tab:capability-mismatch} summarizes recurrent capability mismatches and the engineering controls they motivate.
\begin{table}[H]
\centering
\small
\caption{Typical capability mismatches in agentic software engineering.}
\label{tab:capability-mismatch}
\begin{tabularx}{\linewidth}{@{}p{0.29\linewidth}p{0.31\linewidth}X@{}}
\toprule
\textbf{Capability expansion} & \textbf{Typical failure} & \textbf{Required engineering control} \\
\midrule
Generation outruns intent clarification & Fast implementation of undecided semantics & Intent and specification \\
Local implementation outruns integration & Locally correct changes conflict system-wide & Semantic integration \\
Delegation outruns verification & Candidate changes accumulate & Verification backpressure \\
Local context updates outrun shared-state synchronization & Agents act on stale engineering facts & Context coherence \\
Tool capability exceeds task authority & Agent action radius becomes excessive & Task contract and sandboxing \\
Release outruns responsibility acceptance & Deployment precedes risk closure & Responsibility gate \\
Generation cost falls faster than lifecycle cost & Rework and long-term responsibility obligations are underestimated & Accepted-TC accounting \\
\bottomrule
\end{tabularx}
\end{table}

These failures share one structure: execution has completed while the engineering conditions of the software change remain open.

The order-state example is illustrative. A backend agent may implement \texttt{expired} while a client agent updates the SDK in parallel. If they rely on different versions of the state semantics, both candidates can be locally correct and still compose into a system-level error.

\subsection{Software Change as the Control Object}
Traditional processes manage different stages through requirements, tasks, commits, pull requests, and releases. These objects remain valuable, but none individually spans a complete agentic change.

One software change may cross multiple agents, tasks, and commits while also changing specifications, data contracts, configuration, prompts or instructions, and runtime policy. Engineering governance needs to trace why the change exists, who authorized it, what it altered, how it was verified, how it entered the baseline, what happened in operation, and who accepted the residual risk.

We therefore elevate \emph{software change} to the lifecycle-level control object and formalize it in Section~\ref{sec:tc} as Trustworthy Change.

\paragraph{Five-loop governance.}
The process around TC is
\[
\text{Intent}\rightarrow\text{Delegation}\rightarrow\text{Verification}\rightarrow\text{Integration}\rightarrow\text{Operation}.
\]

\begin{itemize}
\item \textbf{Intent} decides the change goal, success conditions, and prohibited outcomes.
\item \textbf{Delegation} defines execution subjects, permissions, scope, and resources.
\item \textbf{Verification} relates engineering claims to evidence.
\item \textbf{Integration} determines whether the change may enter the authoritative baseline.
\item \textbf{Operation} continues to test the assumptions on which acceptance depended.
\end{itemize}

Specification is the refinement through which Intent becomes delegable. Construction is execution after Delegation. Evidence is the primary output of Verification. Responsibility cuts across all five loops.

\subsection{Repositioning Established Software Engineering Knowledge}
Agent SE does not require reinvention of requirements engineering, architecture, testing, configuration management, formal methods, DevOps, or project management. Their roles become sharper under scalable digital execution.

Requirements engineering constrains semantic freedom; architecture controls change radius and context cost; testing and formal methods construct evidence; configuration management preserves the Candidate--Baseline boundary; DevOps feeds runtime facts back into TC; and project management must additionally govern HACs, agent resources, verification queues, and responsibility acceptance.

TC provides a single object around which these established methods can reconnect.

\section{Execution Units and Responsibility Topology}
\label{sec:topology}

\subsection{Human--Agent Cell: Definition and Interface}
In an agentic environment, the word \emph{developer} no longer fully describes the execution unit that actually performs software work. We define
\[
HAC_i=\langle H_i,A_i,M_i,T_i,P_i,B_i\rangle,
\]
where
\begin{itemize}
\item $H_i$ is the human execution subject;
\item $A_i$ is the set of personal agents;
\item $M_i$ is local working memory or context;
\item $T_i$ is the available tool set;
\item $P_i$ is delegated permission; and
\item $B_i$ is the resource budget.
\end{itemize}

A Human--Agent Cell centers on one human execution subject and combines agents, local context, tools, permissions, and budget into the practical execution unit of software work.

\begin{figure}[H]
\centering
\includegraphics[width=0.94\linewidth]{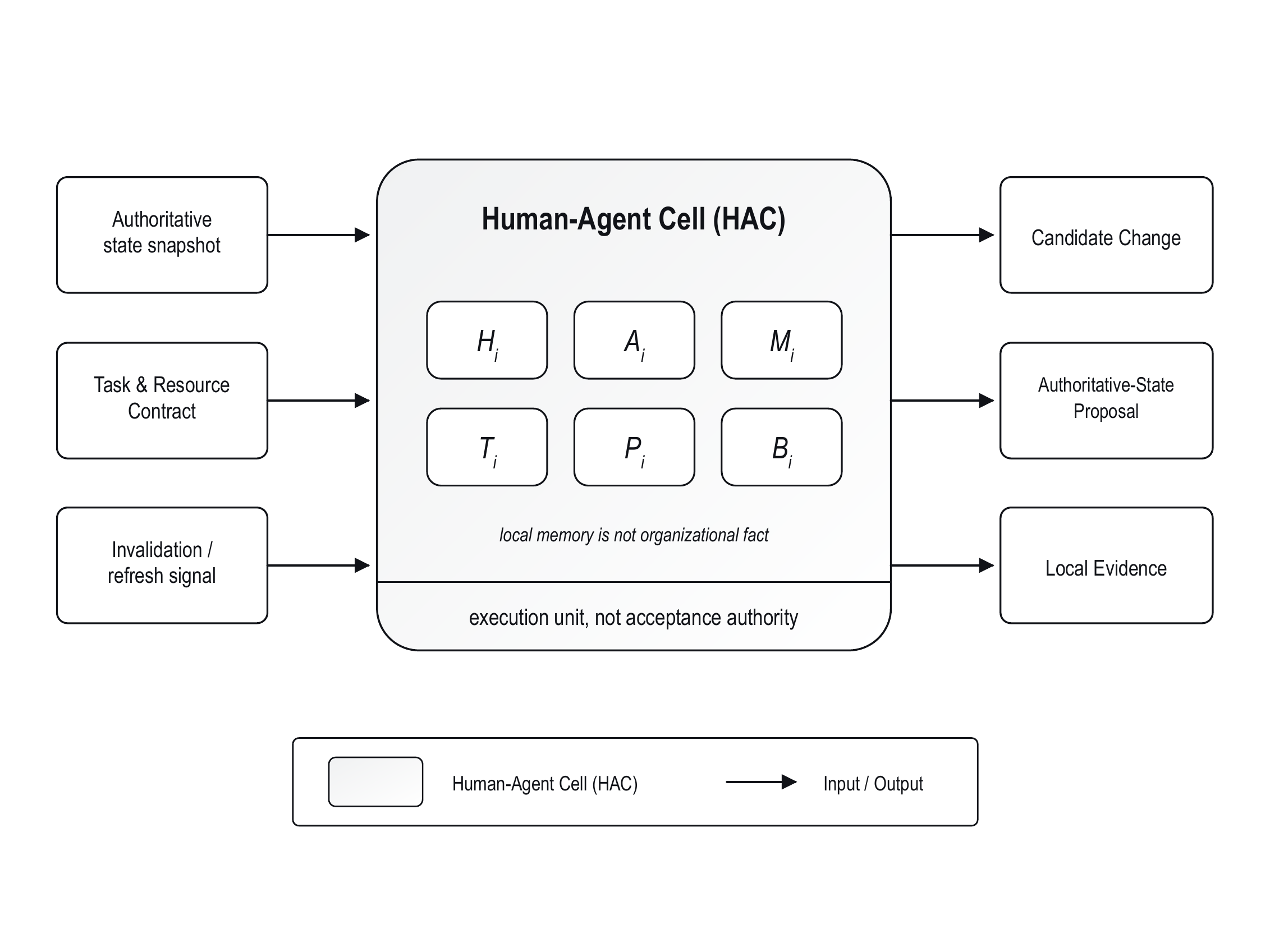}
\caption{A HAC as an execution boundary. It consumes an authoritative-state snapshot, a task/resource contract, and state-change signals, and produces candidates, local evidence, and proposals. Local memory does not automatically become organizational fact.}
\label{fig:hac-boundary}
\end{figure}

A HAC may use one agent or run multiple specialized agents concurrently. Changing a model version, adding agents, or changing a token budget does not usually create a new HAC. A material change in the human execution subject or in the formal engineering boundary is instead a HAC-level handoff, termination, or redefinition. HAC is the human-centered execution boundary studied in this paper; it is not claimed to be the only possible execution unit in an agentic software organization. Team-level, organization-level, or fully autonomous agents may exist outside a HAC while remaining subject to the same distinction between execution authority and residual-risk acceptance authority. HAC separation denotes an execution-boundary distinction, not epistemic independence; different HACs may still share models, tools, retrieval sources, or other common-mode dependencies. Figure~\ref{fig:hac-boundary} summarizes this human-centered execution boundary and its formal inputs and outputs.

\paragraph{Formal interface.}
A HAC can retain substantial local information: agent scratch memory, provisional analyses, debugging hypotheses, unaccepted designs, and candidate alternatives. Such information is execution state rather than organizational truth.

We define the minimal formal input as
\[
\operatorname{Input}(HAC)=\{\text{Authoritative State Snapshot},\text{Task--Resource Contract},\text{State-Change Signal}\},
\]
and its formal output as
\[
\operatorname{Output}(HAC)=\{\text{Candidate Change},\text{Authoritative-State Proposal},\text{Local Evidence}\}.
\]

This boundary permits substantial internal autonomy while constraining state-changing outputs to candidate changes, proposals, and evidence. Organizational fact formation follows
\[
\text{Proposal}\rightarrow\text{Authorized Acceptance}\rightarrow\text{Versioned Commit}.
\]
Local memory and Authoritative Engineering State are therefore distinct by construction.

\subsection{Responsibility Anchor and Effective Governance Rules}
A \textbf{Responsibility Anchor} is
\begin{quote}
A responsibility node traceable to a natural person or an institutionalized organizational role that has formal authority to accept or reject residual risk for a software baseline or a key responsibility domain.
\end{quote}

Examples include a Product Owner, Security Owner, Architecture Owner, or Release Authority, provided that the role actually carries the relevant acceptance right. An agent may receive operational authority inside a delegated boundary and may be recorded as an auditable executor. In the governance model of this paper, acceptance of residual risk remains assigned to an identifiable human or institutional legal subject. This assumption is consistent with current governance frameworks that allocate provider, deployer, and human-oversight duties to natural persons, legal persons, public authorities, or other formal organizational actors~\cite{ref32}. We do not prejudge whether future legal systems may assign legal personality to AI systems.

HACs and Responsibility Anchors belong to different analytical levels:
\[
N_{\mathrm{HAC}}\neq N_{\mathrm{anchor}}.
\]
The same person can be both a HAC execution subject and an anchor for a responsibility domain; a person can also execute under the authority of a different anchor. Table~\ref{tab:agent-hac-anchor} distinguishes the execution and acceptance roles of agents, HACs, and Responsibility Anchors.

\begin{table}[H]
\centering
\small
\caption{Agents, HACs, and Responsibility Anchors.}
\label{tab:agent-hac-anchor}
\begin{tabularx}{\linewidth}{@{}>{\raggedright\arraybackslash}p{0.22\linewidth}>{\raggedright\arraybackslash}p{0.22\linewidth}>{\centering\arraybackslash}p{0.17\linewidth}X@{}}
\toprule
\textbf{Object} & \textbf{Primary role} & \textbf{Produces Candidate?} & \textbf{Final acceptance authority} \\
\midrule
Agent & Digital execution & Yes & No \\
HAC & Local software execution & Yes & Only if its human subject holds the required formal authority \\
Responsibility Anchor & Residual-risk acceptance or rejection & May participate & Yes \\
Team/Organization Agent & Cross-HAC or organizational execution & Yes & No \\
\bottomrule
\end{tabularx}
\end{table}

\subsection{Responsibility Topology}
Responsibility Topology describes how residual-risk acceptance authority is distributed under the currently effective software-governance rules; it is distinct from a generic organizational chart, role matrix, or risk register because the classification depends specifically on independent final acceptance rights for a software baseline or affected responsibility domain.\footnote{The phrase \emph{responsibility topology} also appears in a February 2026 MARIA OS design note, where it denotes a weighted directed graph of decision nodes and responsibility flows with continuous human/agent responsibility allocation; see \url{https://os.maria-code.ai/en/blog/agentic-company-structural-design}. Our use is deliberately narrower: it refers specifically to the topology induced by \emph{independent residual-risk acceptance authority} for software baselines or affected responsibility domains. We make the lexical overlap explicit because the two constructs are adjacent but not equivalent.}

\paragraph{Single-center topology.}
If one final responsibility anchor $R^*$ can (1) finally accept or reject residual risk for the current software baseline and (2) revoke or override the software-engineering authority of other participants without changing the governance rules themselves, the organization has a \textbf{single-responsibility-center} topology. We use \textbf{single-center SE} for this theoretical form. A one-person company (OPC) is a common practical instantiation, but the topology is not defined by headcount.

Single-center SE can contain many HACs, agents, contractors, or external specialists. For a change $x$,
\[
\mathrm{Accept}_{\mathrm{single}}(x)=\mathrm{Accept}_{R^*}(x).
\]

\paragraph{Multi-anchor topology.}
If a software baseline contains at least two independent Responsibility Anchors, each holding acceptance rights in a key domain that another anchor cannot unilaterally revoke under normal governance, the organization has a \textbf{multi-anchor} topology.

Responsibility Topology is an organization- or baseline-level property under the currently effective governance rules. For a particular change $x$ evaluated against authoritative state $S_v$, define the \textbf{effective acceptance set}
\[
K(x,S_v)
\]
as the Responsibility Anchors whose acceptance is mandatory for that change under the current governance rules and authoritative state. When $S_v$ is fixed, we write $K$ for brevity. Then
\[
\mathrm{Accept}_{\mathrm{multi}}(x)=\mathrm{Accept}_{K}(x)
=\IndexedAnd{k\in K}\mathrm{Accept}_{k}(x).
\]
System-level acceptance requires the relevant authority domains to close jointly. A multi-anchor organization may therefore have a particular UI-only TC with $K(x,S_v)=\{R_{\mathrm{Product}}\}$; the organization remains multi-anchor because topology describes the wider distribution of independent acceptance rights, not the cardinality of the effective acceptance set for every change. Quorum, alternative-approver, conditional-acceptance, and emergency-override structures are extensions beyond the present two-topology model.

\paragraph{Operationalizing current governance rules.}
Responsibility Topology cannot be inferred solely from job titles, repository write access, or who happens to merge most changes. Empirical identification should observe at least three sources of evidence.

First, \textbf{de jure governance}: job definitions, approval matrices, contracts, policies, formal ownership, and explicit risk-acceptance rules. Second, \textbf{technical enforcement}: CODEOWNERS, branch protection, IAM, deployment permissions, rollback authority, and release gates. Third, \textbf{de facto practice}: who actually accepts, rejects, revokes, or overrides a change during normal engineering governance.

Technical permissions are not identical to formal risk-acceptance authority. A mismatch among de jure authority, technical enforcement, and de facto practice should be recorded rather than collapsed. While such a conflict remains unresolved, a valid acceptance path has not been established and the candidate cannot be treated as cleanly Eligible. A topology change occurs when management changes the governance rules so that the distribution of anchor authority changes.

\paragraph{Operational distinction between anchor validity and acceptance.}
The R6/Pending distinction becomes operational only if an engineering record can separate two questions. Throughout this paper, $\mathrm{ValidAnchor}(k,x)$, $\mathrm{Accept}_k(x)$, $\mathrm{Eligible}(x,S_v)$, and $\mathrm{AcceptedTC}(x)$ are Boolean-valued predicates over the recorded engineering and governance state rather than quantitative functions. In particular, $\mathrm{Accept}_k(x)=\mathrm{true}$ means that anchor $k$ has formally recorded acceptance of the residual risk of change $x$ under the currently effective governance rules. For a required domain $k$ and change $x$, let $\mathrm{ValidAnchor}(k,x)$ denote whether current governance establishes a legitimate anchor for that domain; let $\mathrm{Accept}_k(x)$ denote the change-specific acceptance decision recorded by that anchor. The first is inferred from de jure rules, technical enforcement, and de facto governance evidence; the second is a state of the particular TC. Thus an absent or invalid required anchor can trigger R6, whereas
\[
\mathrm{ValidAnchor}(k,x)=\mathrm{true},\qquad \mathrm{Accept}_k(x)=\mathrm{false}
\]
is an acceptance-pending condition rather than an R6 failure. These symbols are operational notation, not additional top-level constructs. In practice, CODEOWNERS or deployment permissions can contribute evidence about the first question, but they are not by themselves proof of formal residual-risk authority.

\paragraph{Worked closure example.}
The order-status change makes the topology distinction concrete. In a single-center organization, the required set is
\[
K=\{R^*\},\qquad \mathrm{Accept}_K(x)=\mathrm{Accept}_{R^*}(x).
\]
If the candidate has passed R1--R6 and the final anchor accepts it,
\[
\mathrm{Eligible}(x)=\mathrm{true},\qquad
\mathrm{Accept}_{R^*}(x)=\mathrm{true},
\]
so
\[
\mathrm{AcceptedTC}(x)=\mathrm{true}.
\]
The same artifact change can require a different closure under multi-anchor governance. Suppose Order semantics and Payment timeout behavior belong to independently governed domains:
\[
K=\{R_{\mathrm{Order}},R_{\mathrm{Payment}}\},
\]
\[
\mathrm{Accept}_K(x)=
\mathrm{Accept}_{\mathrm{Order}}(x)
\land
\mathrm{Accept}_{\mathrm{Payment}}(x).
\]
If Order has accepted while Payment has not,
\[
\mathrm{Eligible}(x)=\mathrm{true},\quad
\mathrm{Accept}_{\mathrm{Order}}(x)=\mathrm{true},\quad
\mathrm{Accept}_{\mathrm{Payment}}(x)=\mathrm{false},
\]
the change is \emph{Eligible / Acceptance Pending}, not Accepted. If instead
\[
\mathrm{ValidAnchor}(\mathrm{Payment},x)=\mathrm{false},
\]
R6 fails because the required responsibility path is invalid. Extra reviewers, contractors, or agents change execution or evidence capacity; they do not create a new responsibility center unless governance creates a new independent acceptance domain.

This criterion has two useful invariance properties. First, the topology can remain unchanged while execution capacity changes dramatically. A founder who moves from one coding agent to twenty concurrent agents has changed the execution graph, but not the location of final residual-risk acceptance if the same final anchor retains revocable authority. Likewise, adding contractors, reviewers, or specialist HACs does not by itself create a new responsibility center. Second, the topology can change while headcount remains nearly constant. Two engineers may constitute a multi-anchor organization if each has an independent, non-revocable acceptance right over an affected domain. These counterexamples make Responsibility Topology orthogonal to human count, agent count, and HAC count rather than a proxy for any of them.

The distinction matters only if it changes engineering behavior. Under a single-center topology, conflicting evidence and local interpretations can ultimately be resolved by one final anchor without changing the governance rules. Under a multi-anchor topology, no single HAC or anchor can unilaterally close all affected responsibility domains. The organization must therefore represent which facts are authoritative, which changes invalidate other work, which anchors are required for a particular TC, and whether execution has completed without responsibility closure. In this sense, Responsibility Topology is not merely a descriptive taxonomy: it predicts different coordination and acceptance mechanisms.

\begin{figure}[H]
\centering
\includegraphics[width=0.98\linewidth]{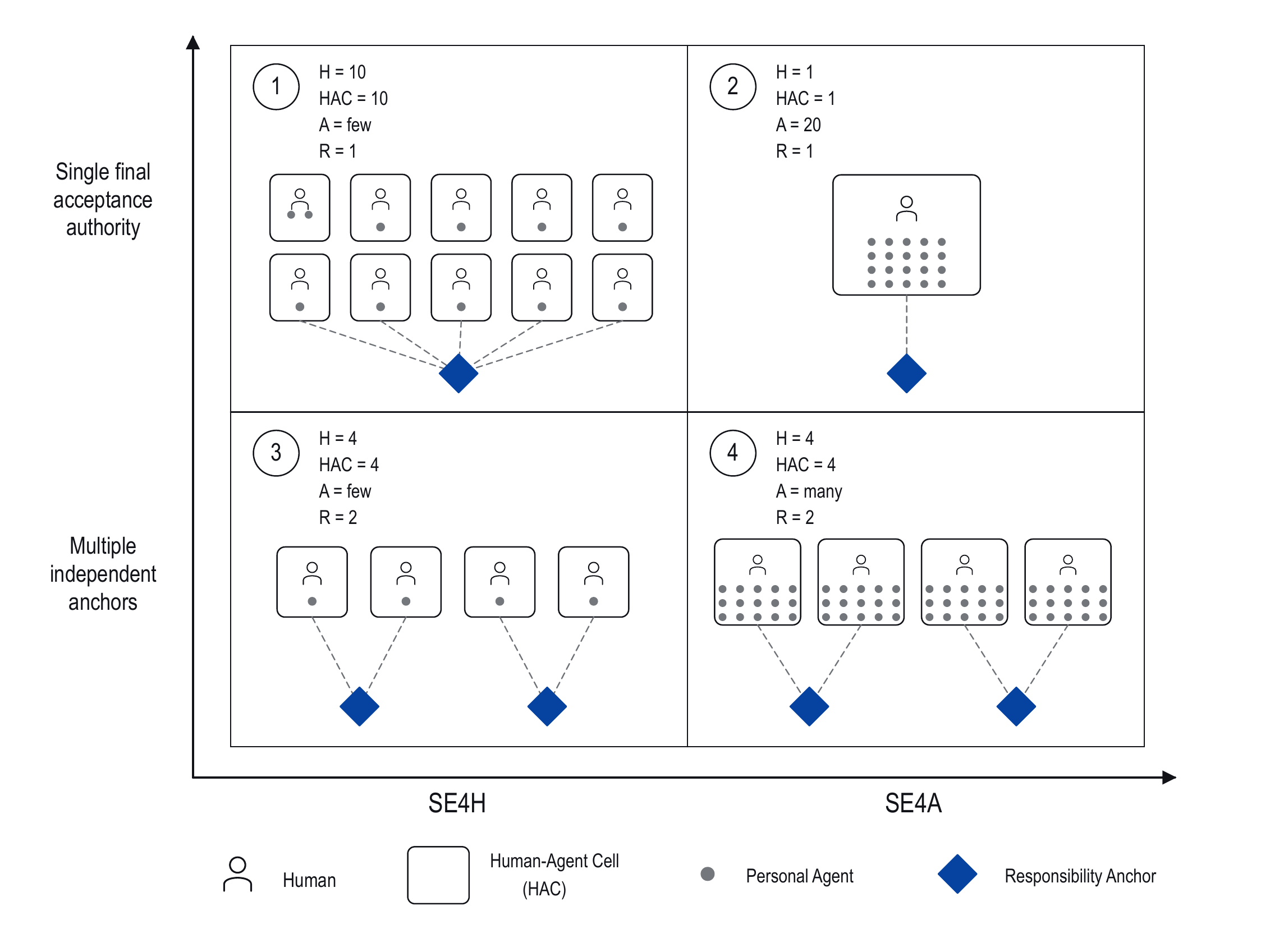}
\caption{Responsibility Topology is orthogonal to SE4H/SE4A-oriented engineering modes. Humans, HACs, personal agents, and independent Responsibility Anchors are distinct quantities.}
\label{fig:orthogonality}
\end{figure}

The boundary is intentionally based on \emph{residual-risk acceptance} rather than ordinary technical permissions. A repository administrator may be able to merge code without possessing formal authority to accept the business or safety risk of the resulting baseline; conversely, a product or compliance role may possess formal acceptance authority without holding direct deployment credentials. Current effective governance rules, evidenced jointly by de jure authority, technical enforcement, and de facto practice, determine the topology. As specified above, unresolved disagreement among these sources leaves the acceptance path invalid for the current TC. Figure~\ref{fig:orthogonality} summarizes the resulting independence of Responsibility Topology from human, HAC, and agent counts.

\subsection{Execution Distribution versus Responsibility Distribution}
Agentic software organizations vary along at least two independent structural dimensions. \emph{Execution distribution} concerns how work is spread across HACs and agents; \emph{responsibility distribution} concerns how independent residual-risk acceptance authority is organized. Context divergence and invalidation arise primarily from distributed execution, whereas joint acceptance and responsibility closure follow from distributed authority. Multi-anchor governance makes explicit authority, versioning, and invalidation semantics necessary when cross-anchor closure depends on shared engineering facts. Table~\ref{tab:execution-vs-responsibility} separates the primary structural sources of the resulting coordination and closure mechanisms.

\begin{table}[H]
\centering
\scriptsize
\caption{Primary structural source of coordination and closure mechanisms. ``Strengthens'' means that multi-anchor governance turns a coordination concern into an explicit cross-anchor closure requirement rather than creating the concern by itself.}
\label{tab:execution-vs-responsibility}
\begin{tabularx}{\linewidth}{@{}>{\raggedright\arraybackslash}p{0.31\linewidth}>{\raggedright\arraybackslash}p{0.29\linewidth}X@{}}
\toprule
\textbf{Mechanism or pressure} & \textbf{Distributed HAC execution} & \textbf{Independent acceptance authority} \\
\midrule
Candidate parallelism & Primary source & Not required \\
Context divergence / staleness & Primary source & Not required \\
Authoritative, versioned state & Useful for coordination & Strengthens: required for cross-anchor closure \\
Context Invalidation & Primary trigger & Strengthens when stale assumptions affect closure \\
Cross-cell verification & Enables heterogeneous evidence sources & May impose domain-specific evidence obligations \\
Joint acceptance & Not required & Primary source \\
Acceptance Pending & Not required & Primary source \\
Responsibility Closure / Orphan & Not required & Primary source \\
Delegation--Responsibility misalignment & Creates opportunities for mismatch & Makes unresolved mismatch non-collapsible by one center \\
\bottomrule
\end{tabularx}
\end{table}

\paragraph{Local memory versus organizational truth.}
Different HACs naturally possess different contexts:
\[
\mathrm{Context}_i(t)\neq\mathrm{Context}_j(t).
\]
If agents act on different versions of a specification, architecture decision, or contract, their candidates can each be locally coherent while being mutually incompatible with the current system semantics.

\paragraph{Delegation and responsibility are different structures.}
The structure that delegates execution is not the structure that accepts risk:
\[
\text{Delegation Graph}\neq\text{Responsibility Graph}.
\]
Agents and HACs can complete their tasks while the anchor responsible for the affected risk domain still rejects the resulting change.

\paragraph{Heterogeneous evidence.}
The value of a team does not come only from additional execution capacity. Different expertise, contexts, interests, and acceptance rights can create heterogeneous judgment:
\[
\text{Heterogeneous Judgment}+\text{Independent Evidence}+\text{Distributed Acceptance}.
\]
Section~\ref{sec:team} develops this mechanism further.

\paragraph{Toward human--agent team software processes.}
PSP emphasizes planning, measurement, and quality discipline in the individual software process; TSP extends this concern to team goals, roles, quality, and collective commitment~\cite{ref4,ref5}. A related separation appears in Agent SE: single-center SE studies how one final responsibility center governs scalable digital execution, while multi-anchor Team SE studies how multiple HACs and independent anchors maintain shared engineering state, cross-cell verification, and system-level responsibility closure. These observations motivate a future research direction on human--agent team software processes; this paper does not define a new fixed process standard.

\section{Trustworthy Change: A Unified Object for Software Change}
\label{sec:tc}

\subsection{TC Components and States}
Prompts, tasks, commits, and pull requests each serve different software-engineering functions. A prompt expresses interaction intent, a task organizes execution, a commit records a versioned code change, and a pull request supports review and integration.

An agentic software change often crosses several of these objects and may additionally modify specifications, data contracts, configuration, model instructions, and runtime policy. If governance is attached only to one local artifact, it becomes difficult to reconstruct the complete relationships among production, verification, configuration, and responsibility.

We define the engineering object that spans the full lifecycle of a software change as \textbf{Trustworthy Change (TC)}.

The need for a lifecycle-spanning object follows from a mismatch among existing artifacts. A commit captures code history but not necessarily the intent that justifies the change. A pull request can aggregate discussion and tests but may not encode the resource envelope, runtime assumptions, or authority structure under which the change is accepted. A change request captures organizational intent but may be detached from the concrete provenance and evidence of the implementation. A release identifies a delivered baseline but is too late to serve as the unit that structures delegation and verification. Agentic execution makes these gaps more visible because one semantic change can be spread across many agent trajectories, commits, tool calls, generated tests, and state proposals.

TC acts as a \emph{cross-artifact identity} for one consequential software change. Its six components are not intended as six new documents. They form the compact working set of traceable closure dimensions used in this framework to answer six engineering questions: what was intended, under what authority and resources execution occurred, what changed and where it came from, what evidence supports the claims, how the change enters and survives operation, and who can accept the remaining risk. We do not claim that this decomposition is ontologically exhaustive or uniquely minimal. Omitting one of these dimensions exposes a corresponding class of governance or assurance failure that the framework is designed to make visible: ambiguous intent, unbounded action, untraceable provenance, unsupported claims, unsafe integration, or unclosed responsibility.

\paragraph{Components.}
Table~\ref{tab:tc-components} summarizes the six working closure dimensions used by TC.
\begin{table}[H]
\centering
\small
\caption{Six components of Trustworthy Change.}
\label{tab:tc-components}
\begin{tabularx}{\linewidth}{@{}p{0.31\linewidth}X@{}}
\toprule
\textbf{Component} & \textbf{Main content} \\
\midrule
Intent \& Specification & Purpose, success conditions, failure conditions, and prohibited outcomes \\
Delegation \& Resource Envelope & Execution subjects, permissions, scope, and resources \\
Change \& Provenance & The actual change, its origin, dependencies, and execution history \\
Evidence & Engineering claims and the evidence supporting them \\
Integration \& Runtime State & Baseline, release, operation, observation, and recovery state \\
Responsibility & Proposal, authorization, verification, acceptance, and residual-risk ownership \\
\bottomrule
\end{tabularx}
\end{table}

TC can simultaneously serve as a production unit, verification unit, configuration/change-management unit, measurement unit, economic-accounting unit, and responsibility unit. This unification does not require all engineering information to be stored in one physical record. It requires that the information remain traceable to the same underlying software change.

\paragraph{State progression.}
A HAC first produces a \textbf{Candidate Change}. Candidate means digital execution has produced an artifact ready for engineering evaluation. Whether or not the candidate is eventually accepted, token, compute, tool, and human-attention resources have already been consumed and therefore belong in the engineering cost.

A Candidate becomes \textbf{Eligible} after passing engineering-qualification checks:
\[
\mathrm{Eligible}(x)\iff\neg\left(R_1\lor R_2\lor R_3\lor R_4\lor R_5\lor R_6\right).
\]
Eligible means the change is qualified to enter formal responsibility acceptance.

For a change $x$ evaluated against authoritative state $S_v$, let $K=K(x,S_v)$ denote the effective acceptance set defined in Section~\ref{sec:topology}. System-level acceptance is
\[
\mathrm{Accept}_K(x)=\IndexedAnd{k\in K}\mathrm{Accept}_k(x),
\]
and therefore
\[
\mathrm{AcceptedTC}(x)\iff \mathrm{Eligible}(x)\land \mathrm{Accept}_K(x).
\]

Eligibility and acceptance answer different questions: the former is a qualification judgment; the latter is formal closure of residual risk.

This separation is especially important when generation is faster than organizational decision making. A candidate can be technically well formed and richly evidenced while still waiting for a required anchor; conversely, organizational willingness to proceed cannot make an inadequately evidenced candidate Eligible. The three-state progression prevents implementation completion, engineering qualification, and responsibility acceptance from collapsing into a single overloaded notion of ``done.'' It also creates a stable accounting boundary: all candidate work consumes resources, only Eligible changes may enter formal acceptance, and only Accepted TC enters the authoritative baseline.

Eligibility is relative to the current governance rules, the current Authoritative-State version, and the currently available evidence. Tools and people can jointly determine it. R2, R3, and parts of R5 can often be highly automated; R1 and R4 frequently require semantic or professional judgment; R6 requires analysis of formal authority.

In particular,
\[
\mathrm{Eligible}(x,S_v)\not\Rightarrow\mathrm{Eligible}(x,S_{v+1}).
\]
If an authoritative fact that affects $x$ changes, eligibility must be revalidated through the Context Invalidation actions defined in Section~\ref{sec:team}.

\begin{figure}[H]
\centering
\includegraphics[width=0.98\linewidth]{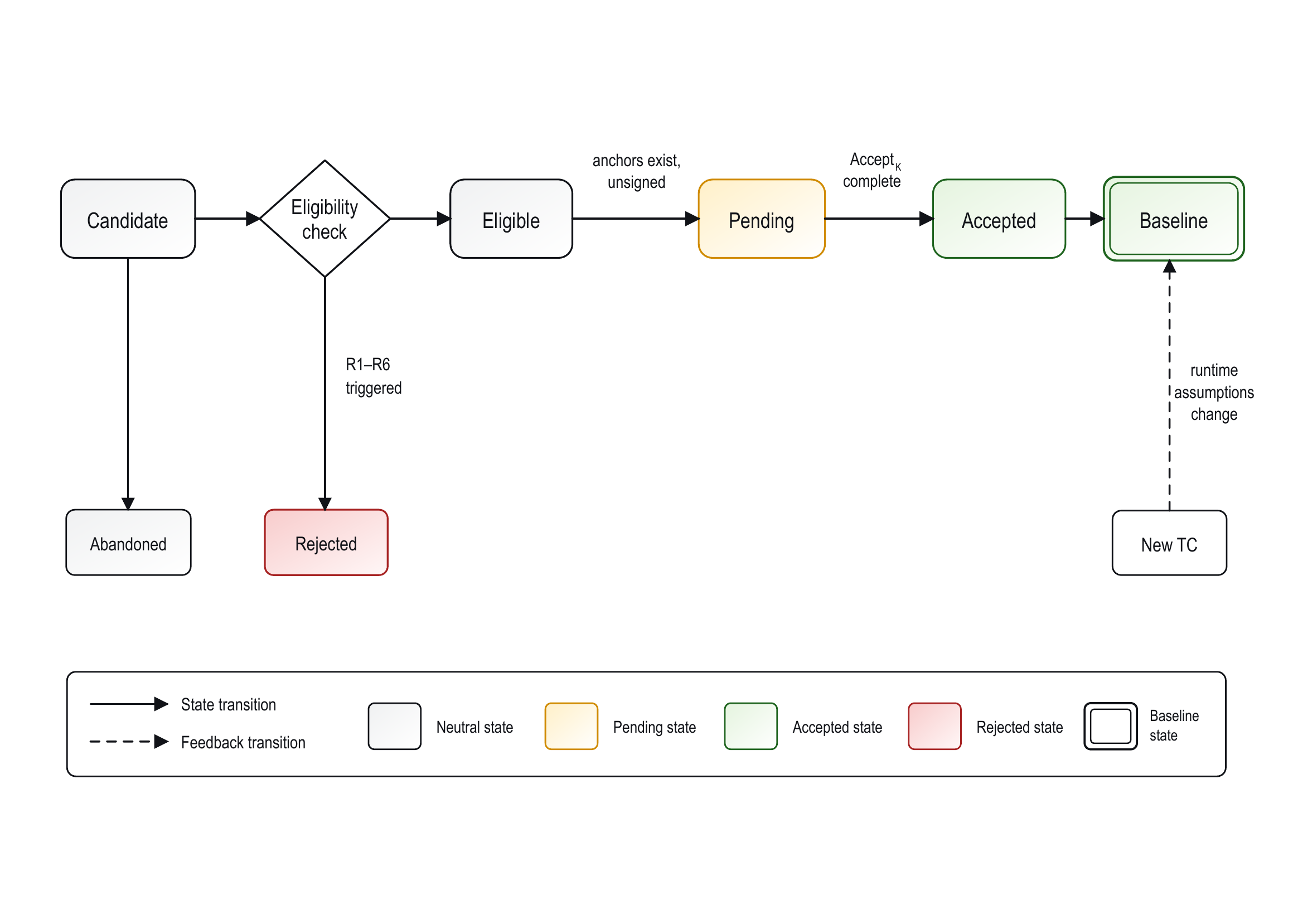}
\caption{Trustworthy Change state machine. Eligibility denotes engineering qualification; acceptance denotes closure of residual risk. A valid but unsigned anchor produces an acceptance-pending condition rather than an R6 failure.}
\label{fig:tc-state-machine}
\end{figure}

If the required anchors validly exist but have not yet accepted the change, then
\[
\mathrm{Eligible}(x)=\mathrm{true},\qquad \mathrm{Accept}_K(x)=\mathrm{false}.
\]
The change is \textbf{Eligible / Acceptance Pending}; this is not an R6 failure. An agent that produced the candidate may assist with eligibility analysis, but it cannot by itself close the responsibility structure required for its own candidate. Figure~\ref{fig:tc-state-machine} summarizes the resulting qualification and acceptance states.

\subsection{Eligibility, Rejection, and Acceptance}
Table~\ref{tab:r1-r6} lists the conditions that prevent a Candidate from becoming Eligible.
\begin{table}[H]
\centering
\small
\caption{Conditions that prevent a Candidate from becoming Eligible.}
\label{tab:r1-r6}
\begin{tabularx}{\linewidth}{@{}p{0.08\linewidth}p{0.30\linewidth}X@{}}
\toprule
\textbf{ID} & \textbf{Failure type} & \textbf{Typical issue} \\
\midrule
R1 & Intent / Specification Failure & Goal, non-goals, boundary, or key semantics remain undecided \\
R2 & Delegation / Resource Failure & Authority is exceeded, scope is wrong, or resource budgets are uncontrolled \\
R3 & Change / Provenance Failure & Origin, dependencies, or execution process cannot be traced \\
R4 & Evidence Failure & A material claim lacks sufficient evidence \\
R5 & Integration / Operation Failure & Baseline, release, observability, or recovery readiness is insufficient \\
R6 & Responsibility Failure & A required responsible subject or acceptance path does not exist, is invalid, or lacks legitimate authority \\
\bottomrule
\end{tabularx}
\end{table}

R1 and R6 are related but logically independent. R1 asks whether intent and specification are sufficiently decided. R6 asks whether the responsibility structure under which those decisions and subsequent risk acceptance occur exists and is valid.

A legitimate Product Anchor may already exist while the requirement remains ambiguous; R1 fails while R6 does not. Conversely, the specification may be fully clear while no valid anchor exists for the affected responsibility domain; R6 fails while R1 does not. Semantic commitments relevant to R1 must be made by subjects authorized under the Responsibility Topology, but R1 cannot be replaced by R6.

For a pre-production candidate, R5 evaluates operational readiness---for example release strategy, observability, and recovery planning---rather than requiring proof that future operation can never fail.

Trustworthiness is not a permanent label. An Accepted TC is accepted relative to a particular version, evidence set, and time condition. New runtime facts can invalidate the acceptance assumptions and trigger a new TC.

\subsection{Lifecycle and Responsibility}
The basic TC process remains
\[
\text{Intent}\rightarrow\text{Delegation}\rightarrow\text{Verification}\rightarrow\text{Integration}\rightarrow\text{Operation}.
\]
Intent creates goals and constraints; Delegation creates tasks, permissions, and resource envelopes; Verification relates claims to evidence; Integration prepares and validates the composition of an Eligible Change with the target baseline; Operation produces new runtime evidence. Final baseline commitment occurs only after the required acceptance set $K$ closes:
\[
\text{Integration Readiness}\rightarrow\mathrm{Accept}_K\rightarrow\text{Baseline Commit}\rightarrow\text{Operation}.
\]

Specification is progressively refined between Intent and Delegation. Construction is execution after Delegation. Responsibility spans all five loops.

\paragraph{Responsibility checkpoints.}
Each loop corresponds to a responsibility question:
\begin{itemize}
\item \textbf{Intent:} who has authority to define or change the software goal?
\item \textbf{Delegation:} who has authority to authorize HAC and agent action?
\item \textbf{Verification:} who judges that the evidence is sufficient for the current claims?
\item \textbf{Integration:} who may admit the change to the authoritative baseline?
\item \textbf{Operation:} who accepts runtime residual risk and recovery obligations?
\end{itemize}
Responsibility is therefore a longitudinal property of TC, not a sixth stage added to the software process.

\paragraph{Three analytical scales of responsibility.}
We use responsibility at three distinct scales. First, the responsibility component inside a TC records who proposed, authorized, verified, and accepted a specific change: an \textbf{instance-level responsibility record}. Second, responsibility checkpoints across the five loops describe the governance questions that must close during the lifecycle: a \textbf{process-level governance concern}. Third, Responsibility Topology describes the formal organization of residual-risk acceptance rights among anchors: an \textbf{organization-level authority structure}. The three levels share a common responsibility semantics, but answer different-granularity questions.

\paragraph{Four assurance questions.}
A TC can be continuously assessed through four questions:

\textbf{Q1. Are intent and specification sufficiently decided?}

\textbf{Q2. Are the claims about the change sufficiently evidenced?}

\textbf{Q3. Does operation continue to support the acceptance assumptions?}

\textbf{Q4. Does responsibility remain identifiable and actionable?}

Security, performance, compatibility, structural correctness, and similar properties enter Q2 as claims supported by corresponding evidence.

\subsection{Relation to Change Management and the Running Example}
TC maps directly onto established change and configuration-management concepts. Table~\ref{tab:tc-change-management} shows the corresponding mapping.

\begin{table}[H]
\centering
\small
\caption{Mapping from traditional change management to Trustworthy Change.}
\label{tab:tc-change-management}
\begin{tabularx}{0.9\linewidth}{@{}p{0.34\linewidth}X@{}}
\toprule
\textbf{Traditional object} & \textbf{Location in TC} \\
\midrule
Change Request & Intent \& Specification \\
Impact Analysis & Specification refinement and risk analysis \\
Work Assignment & Task \& Resource Contract \\
Implementation & Candidate Change \\
Review / Testing & Evidence \\
Configuration Item & Integration \\
Baseline & Accepted state \\
Change Approval & Responsibility acceptance \\
Release / Monitoring & Integration \& Runtime State \\
\bottomrule
\end{tabularx}
\end{table}

Agentic execution adds delegated permission, resource envelopes, agent provenance, evidence independence, runtime drift, and Responsibility Topology. TC can therefore be understood as an extensible engineering abstraction that reorganizes traditional change management for agentic execution conditions.

\paragraph{Running illustrative example.}
The order-state change is the paper's \textbf{running illustrative example}. It is used to maintain conceptual consistency and to show how intent, delegation, evidence, Responsibility Topology, and Context Invalidation apply to the same software change. It is not a case study and does not empirically validate the framework.

After the state space evolves from
\[
\{\texttt{active},\texttt{cancelled}\}
\]
to
\[
\{\texttt{active},\texttt{cancelled},\texttt{expired}\},
\]
Intent and Specification must at minimum define the semantic boundary between \texttt{cancelled} and \texttt{expired}, late payment callbacks, refund states, old-SDK behavior, and historical-data policy.

Delegation constrains the scope, permissions, and budgets of backend, payment, client, and analytics tasks. Verification must cover the state machine, payment contract, late callbacks, legacy-client compatibility, and analytics consistency. Integration must specify contract versions, compatibility windows, release order, rollback, and monitoring. Responsibility is determined by the current Responsibility Topology.

The point of the example is that a small artifact change need not imply a small semantic scope, verification scope, or responsibility scope for the TC.

\section{Common Engineering Methods for Agent Software Engineering}
\label{sec:methods}

This section defines the common method layer once. Sections~\ref{sec:single-center} and~\ref{sec:team} do not introduce second versions of these mechanisms; they discuss how the same specification, task, verification, integration, and recovery methods respond to distributed execution and how their closure conditions differ when acceptance authority is concentrated or distributed.

\subsection{Intent and Progressive Specification}
Natural language can express software intent at low cost, but an agent's ability to adapt to language can also conceal unresolved issues. Intent Engineering therefore begins by distinguishing information with different epistemic and governance status:
\begin{itemize}
\item \textbf{Fact}: a state that can already be observed or verified;
\item \textbf{Evidence}: data or facts supporting a judgment;
\item \textbf{Assumption}: a judgment on which the current analysis depends but that is not yet sufficiently verified;
\item \textbf{Preference}: a stakeholder inclination; and
\item \textbf{Decision}: a formal choice made by a subject with appropriate authority.
\end{itemize}

In the order example, ``the current \texttt{cancelled} state conflates two closure causes'' is a fact. Customer-support and operational data may provide evidence. ``The new enum will not break old clients'' is an assumption until verified. ``Users should be able to distinguish the two closure modes'' is a preference. Formally introducing \texttt{expired} is a decision.

Agents can support feedback clustering, historical retrieval, counterexample construction, and generation of clarification questions. Decisions that change software semantics should enter Authoritative Engineering State.

\paragraph{Progressive specification.}
Natural-language specification creates costs in two directions. Under-specification gives agents a large interpretation space and can increase candidate divergence, clarification effort, and rework. Over-specification increases elicitation, modeling, maintenance, and context costs.

We therefore model specification as a progressive path:
\[
\begin{aligned}
\text{Natural Intent}&\rightarrow\text{Clarification}
\rightarrow\text{Structured Specification}\\
&\rightarrow\text{Executable Constraint}
\rightarrow\text{Formal Specification}.
\end{aligned}
\]

\paragraph{Minimum Sufficient Specification.}
The goal is a \textbf{Minimum Sufficient Specification}:
\begin{quote}
The lower practical boundary of a task- and risk-dependent effective specification region, if such a region exists.
\end{quote}
The hypothesis is not that one universal minimum exists across tasks; rather, additional rigor may eventually yield diminishing reductions in semantic divergence, rework, or verification uncertainty relative to its added lifecycle cost.

Required rigor depends primarily on ambiguity, impact, irreversibility, and verifiability. Agents are well suited to ambiguity detection, generation of clarification questions, structured-specification drafting, generation of candidate contracts/tests/properties, and consistency checking. Semantic commitment remains with the authorized human responsibility subject.

\begin{figure}[H]
\centering
\includegraphics[width=0.93\linewidth]{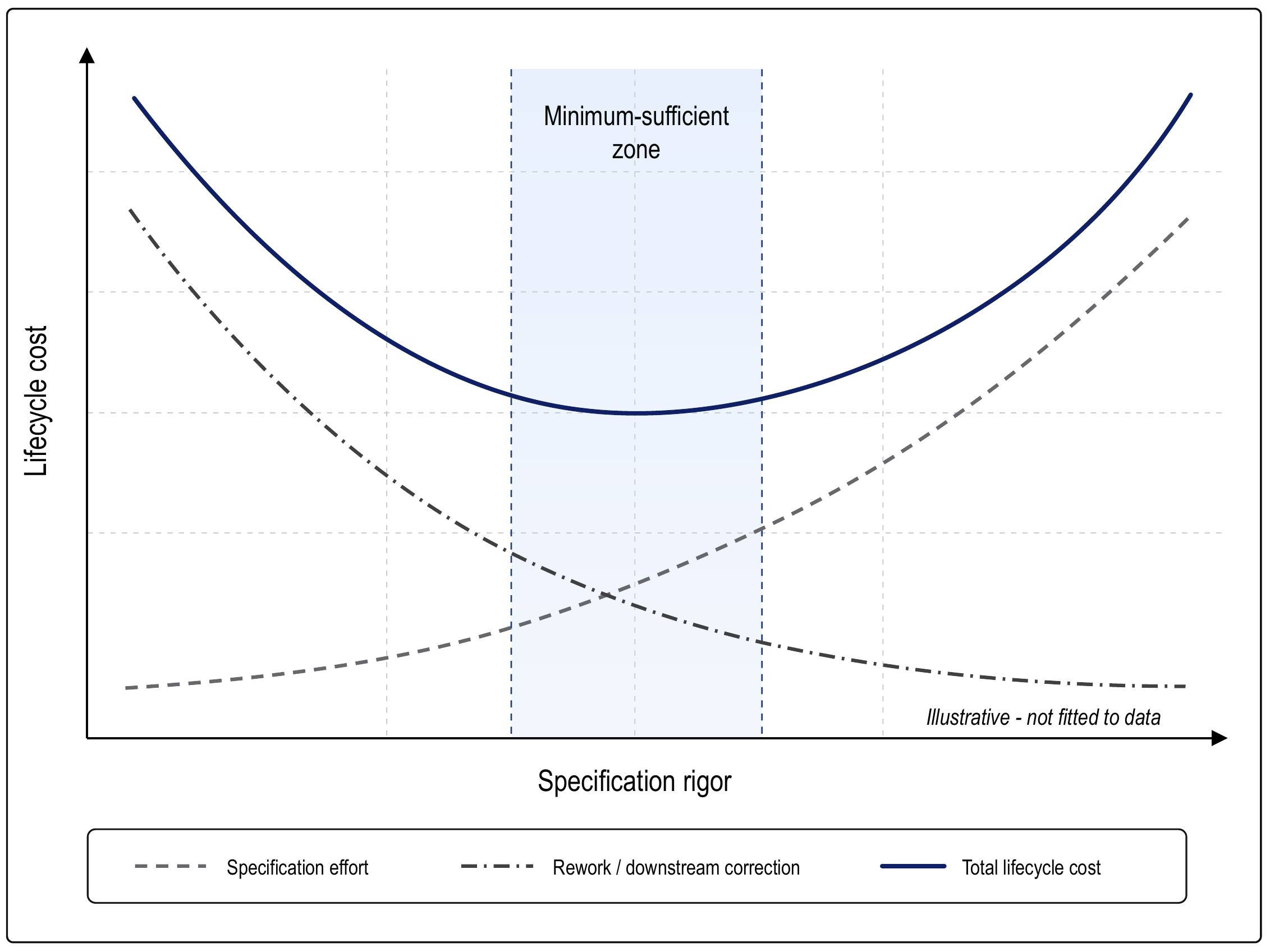}
\caption{Hypothesized lifecycle cost trade-off in Progressive Specification. The curve shapes and the existence of a minimum-sufficient region are empirical hypotheses tested by the research agenda in Section~\ref{sec:agenda}.}
\label{fig:spec-cost}
\end{figure}

The curves in Figure~\ref{fig:spec-cost} are conceptual illustrations: neither their shape nor the existence or location of a minimum-sufficient region is an empirical result of this paper. The research question is whether, for practically relevant classes of changes, there is a level of specification rigor beyond which additional specification cost no longer buys commensurate reductions in downstream ambiguity, rework, or verification uncertainty.

\subsection{Architecture, Change Locality, and Task Contracts}
An architecture intended for agentic software work should satisfy three properties.

\textbf{Agent-operable} means that the system exposes stable commands, clear schemas, repeatable environments, and callable tools so that agents can act reliably.

\textbf{Human-understandable} means that responsible people can still understand key states, dependencies, interfaces, and invariants.

\textbf{Human-takeoverable} means that agent tasks can be stopped, reproduced, rolled back, handed off, and recovered.

We use \emph{change locality} to describe the engineering span of a typical change. It is not defined as an additive scalar. At minimum, it should be observed along three dimensions:
\begin{enumerate}
\item \textbf{Artifact span}: how many modules, files, interfaces, and data structures are touched;
\item \textbf{Context span}: how much engineering context and how many invariants must be loaded for execution and verification; and
\item \textbf{Responsibility span}: how many HACs, responsibility domains, and acceptance anchors are involved.
\end{enumerate}
Review Scope and Revalidation Scope can be observed separately when useful.

Poor change locality is expected to correlate with higher context, integration, and verification cost, but this is an empirical claim rather than a definition. Information hiding and modularity~\cite{ref3} therefore acquire an additional agent-economics interpretation: good modularity may reduce not only human cognitive cost but also agent context and verification cost.

For security, authorization, accounting, and other critical logic, an architecture may use a \textbf{Deterministic Core + Probabilistic Shell}: unacceptable consequences should be bounded by deterministic mechanisms even when probabilistic agents perform surrounding work.

\paragraph{Task and Resource Contract.}
Delegation to an agent must constrain goals, authority, and resources simultaneously. We represent a task as
\[
Task=\{\text{Goal},\text{Context},\text{Scope},\text{Permission},\text{Budget},\text{Acceptance},\text{Stop},\text{Handoff}\},
\]
and the resource envelope as
\[
R=\left\langle\text{Token},\text{Compute},\text{Tool},\text{Wall-Clock Time},\text{Human Attention}\right\rangle.
\]

The Stop Condition is first class. An agent should stop current autonomous execution and escalate when, for example, it needs unauthorized data, expanded write permission, an irreversible migration, budget overrun, or a high-risk consequence that cannot be judged within the current boundary. The design grants agents substantial autonomy inside explicit authorization while keeping the boundary auditable.

\paragraph{Agent execution threat model.}
The central security risk in agentic execution arises from coupling untrusted input to high-privilege tools. OWASP and NIST identify prompt injection, indirect prompt injection, and agent hijacking as major risks in current agent systems~\cite{ref31,ref35}. Peer-reviewed work has begun to connect these threats to action boundaries: Task Shield evaluates whether instructions and tool calls remain aligned with the user's task~\cite{ref39}, while MCPTox studies tool poisoning through malicious MCP metadata and tool descriptions~\cite{ref40}.

In software-engineering settings, attacks may originate in user prompts or in webpages, email, files, RAG content, tool descriptions, and tool outputs read by an agent. Representative threats include:
\begin{itemize}
\item \textbf{Direct Prompt Injection}: user input attempts to override system or task instructions;
\item \textbf{Indirect Prompt Injection}: external documents, pages, email, or retrieved content carry malicious instructions;
\item \textbf{Tool / Context Poisoning}: malicious tool metadata or outputs contaminate agent context;
\item \textbf{Data Exfiltration}: sensitive information is leaked through network access, email, file writes, or tool calls;
\item \textbf{Privilege Escalation / Unauthorized Action}: the agent is induced to perform database writes, IAM changes, external messaging, deployment, or other actions beyond task authority; and
\item \textbf{Persistent Context Poisoning}: malicious content enters persistent memory or shared state and influences later agents.
\end{itemize}

The security objective is not to assume that an agent can recognize every attack. Instead,
\[
\text{Compromise of Interpretation}\not\Rightarrow\text{Unbounded Compromise of Action}.
\]
Even when semantic interpretation is manipulated, permission boundaries, sandboxes, tool hooks, egress policies, and high-impact confirmation should bound the maximum harm radius. This containment-oriented view is also consistent with secure-development guidance such as the NIST Secure Software Development Framework, which treats security practices as lifecycle-integrated engineering controls rather than a final inspection step~\cite{ref7}. Table~\ref{tab:agent-threats} summarizes the principal attack surfaces and controls.

\begin{table}[H]
\centering
\small
\caption{Agent execution threats and engineering controls.}
\label{tab:agent-threats}
\begin{tabularx}{\linewidth}{@{}p{0.27\linewidth}p{0.28\linewidth}X@{}}
\toprule
\textbf{Threat} & \textbf{Primary attack surface} & \textbf{Primary engineering control} \\
\midrule
Direct Prompt Injection & User input & Task scope, instruction/data separation, forbidden actions \\
Indirect Prompt Injection & Web / email / file / RAG & Untrusted-context marking, tool restriction, output validation \\
Tool / MCP Poisoning & Tool description / output & Allowlist, trusted tool registry, runtime validation \\
Data Exfiltration & Network / email / file write & DLP, destination policy, network-egress control \\
Privilege Escalation & DB / IAM / administrative tools & Least privilege, scoped credentials, approval gates \\
Persistent Context Poisoning & Memory / shared state & Proposal/acceptance separation, versioned authoritative state \\
\bottomrule
\end{tabularx}
\end{table}

\subsection{Construction and Evidence-Based Verification}
The engineering objective of agent-native construction is
\begin{quote}
To produce Candidate Changes that are comparable, verifiable, and revocable inside controlled environments.
\end{quote}

Candidate changes should preferably be executed in branches, worktrees, containers, VMs, sandboxes, or equivalent isolated environments. Minimum useful provenance should trace the model or agent, context version, tool calls, dependency changes, manual edits, and build/test environment.

Agents lower the cost of producing alternatives, making multi-candidate exploration more practical. Yet every additional candidate increases comparison and verification cost. Candidate exploration must therefore also be constrained by resource and verification capacity.

\paragraph{Evidence-based verification.}
Verification begins by stating what properties the change claims to satisfy. For the order-state TC, example claims include
\[
C_1:\quad \text{Payment Timeout}\Rightarrow\texttt{expired},
\]
\[
C_2:\quad \text{User Cancellation}\Rightarrow\texttt{cancelled},
\]
as well as claims about late payment, legacy clients, and analytics.

Verification can then be represented as a claims--evidence relation:
\[
\mathrm{Claim}_i\rightarrow\{\mathrm{Evidence}_{i1},\mathrm{Evidence}_{i2},\ldots\}.
\]

Evidence quality depends on both coverage and independence. We distinguish Model/Method Independence, Context Independence, Criterion Independence, and Responsibility Independence.

\paragraph{Common-mode epistemic failure.}
Hallucination or confabulation is one visible source of evidence failure; NIST identifies confident production of erroneous or fabricated content as a generative-AI risk~\cite{ref31}. The broader problem in Agent SE is \textbf{correlated or common-mode epistemic failure}. Agents executed independently may still share a base model, training data, omitted specification assumptions, retrieval sources, benchmark habits, or organizational priors, producing correlated mistakes.

Using different model families can reduce some common failure sources, but model diversity alone does not establish evidence independence. Different models can share data, specifications, retrieval sources, or systematic biases.

\paragraph{Risk-adaptive evidence expectations.}
The meaning of ``sufficiently evidenced'' in Q2 depends on claim type and risk. Table~\ref{tab:risk-evidence} gives illustrative risk-adaptive evidence expectations.

\begin{table}[H]
\centering
\small
\caption{Risk-adaptive evidence expectations.}
\label{tab:risk-evidence}
\begin{tabularx}{\linewidth}{@{}p{0.43\linewidth}X@{}}
\toprule
\textbf{Change / claim type} & \textbf{Illustrative minimum evidence expectation} \\
\midrule
Low risk, local, easily reversible & Deterministic test + provenance + reproducible command \\
Cross-module or public contract & Regression + contract test + executable reproduction \\
Data or state migration & Invariant + representative migration test + rollback rehearsal \\
High-risk concurrency, payment, or authorization logic & Property/fuzz/concurrency test + independent verification \\
High-impact security or compliance property & Independent review + policy check + explicit acceptance \\
Very high risk and formalizable property & Executable/formal constraint + independent evidence \\
\bottomrule
\end{tabularx}
\end{table}

These are not new correctness levels. They are risk-adaptive Evidence Profiles under Q2, shaped by risk, specification rigor, and responsibility domain. Different anchors may declare different minimum evidence gates for their domains.

The approach borrows engineering-quality ideas such as being documented, consistent, complete, exercisable, and reproducible, but we do not call Candidate Change verification ``Artifact Evaluation'' because that term has an established meaning in research-artifact assessment.

\subsection{Integration, Operation, and Recovery}
After local verification, a Candidate must still be integrated at system level. Conventional CI/CD is often represented as
\[
\text{Code}\rightarrow\text{Build}\rightarrow\text{Test}\rightarrow\text{Deploy}.
\]
At the TC level, the lifecycle is richer:
\[
\begin{aligned}
\text{Intent}&\rightarrow\text{Specification}\rightarrow\text{Candidate}\rightarrow\text{Evidence}\\
&\rightarrow\text{Eligible}\rightarrow\mathrm{AcceptedTC}\rightarrow\text{Release}\rightarrow\text{Runtime Evidence}.
\end{aligned}
\]
Only an Accepted TC may modify the authoritative baseline.

Integration must also resolve semantic conflict. Two candidates can merge cleanly in Git while contradicting one another at the level of API contracts, state models, data semantics, or architecture rules. Agent SE must distinguish textual merge from semantic integration.

\paragraph{AgentOps, recovery, and responsibility continuity.}
After release, the system continues to produce technical telemetry, user outcomes, agent behavior, cost, and incident evidence. Agentic operation adds sources of drift including model drift, context drift, cost drift, and architecture drift.

When anomalies occur, autonomy can contract progressively:
\[
\text{Auto-execute}\rightarrow\text{Human approval}\rightarrow\text{Read-only}\rightarrow\text{Stop}.
\]
Systems should also support rollback, restore, reproduce, and handoff. Table~\ref{tab:methods-topology} summarizes how these common methods interact with distributed execution and Responsibility Topology.

\begin{table}[H]
\centering
\small
\caption{How common engineering methods interact with distributed execution and Responsibility Topology.}
\label{tab:methods-topology}
\begin{tabularx}{\linewidth}{@{}p{0.20\linewidth}p{0.35\linewidth}X@{}}
\toprule
\textbf{Method} & \textbf{Single-center SE} & \textbf{Multi-anchor Team SE} \\
\midrule
Intent decision & Converges to one final anchor & May require joint acceptance across responsibility domains \\
Specification & Concurrent HACs may still require context invalidation after authoritative-state changes & The same invalidation problem exists; explicit authority/version semantics support cross-anchor closure \\
Architecture & Controls cognitive burden on one final anchor & Also defines context, ownership, and responsibility boundaries \\
Delegation & Authorizer, verifier, and acceptor may be the same anchor at different times & The three roles may belong to different HACs and anchors \\
Verification & Relies more on tools, heterogeneous models, and external experts & Cross-cell evidence adds organizational heterogeneity \\
Integration & One center closes baseline acceptance & Multiple anchors jointly close responsibility \\
Takeover & Focuses on replacing the final anchor & Requires reconnecting the Responsibility Graph \\
\bottomrule
\end{tabularx}
\end{table}

\section{Single-Center Software Engineering}
\label{sec:single-center}

Section~\ref{sec:methods} defined the common engineering methods. This section focuses on the capacity, verification, continuity, and economic consequences that arise when one final Responsibility Anchor closes residual risk. A one-person company is treated only as a common practical alias for this topology.

\subsection{Capacity, Verification, and Evidence Independence}
A single-center organization can contain
\[
R^*\longrightarrow\{HAC_1,HAC_2,\ldots,HAC_n\}.
\]
The periphery may additionally include security specialists, legal advisers, contract developers, cloud services, and independent testing organizations. These actors can provide execution, consultation, or evidence; formal acceptance of residual risk for the software baseline still closes at $R^*$.

The structural property of single-center SE is \emph{distributed execution capacity with concentrated final responsibility}.

\paragraph{Capacity and the verification queue.}
The order-state change can be delegated simultaneously to backend, payment, SDK, compatibility, and analytics agents. All resulting candidates eventually return to one responsibility center for understanding, verification, and integration.

A conceptual upper bound on effective concurrency is
\[
C_{\mathrm{single}}\le
\min\left(C_D,B_{\mathrm{integration}},B_{\mathrm{verification}},B_{\mathrm{economic}}\right),
\]
where $C_D$ is the decomposability-induced throughput ceiling, $B_{\mathrm{integration}}$ is integration bandwidth, $B_{\mathrm{verification}}$ is verification bandwidth, and $B_{\mathrm{economic}}$ is economic capacity. In practical terms, $C_D$ is the throughput ceiling induced by how far a change can be partitioned into independently verifiable subtasks; changes differ in how much safe parallelism they admit. Under a fixed workload, all terms denote effective throughput ceilings expressed in a common unit such as admissible TCs per unit time. The bound is schematic rather than a calibrated empirical law.

Its practical implication is that adding execution units can increase the arrival rate of candidates without increasing the service rate of the verification and acceptance queues. When candidate supply exceeds those downstream capacities, the organization does not merely become ``busy'': it accumulates partially understood changes, compresses review time, and creates pressure to substitute summaries or agent-generated confidence signals for independent inspection. The resulting bottleneck is structural because the final responsibility center still has to maintain enough understanding to accept residual risk coherently across the baseline.

This also explains why more reviewers do not automatically imply a different topology. Additional HACs can raise verification capacity while the same $R^*$ remains the final acceptance authority. A topology change is justified only when a stable responsibility domain acquires an independent acceptance right that the existing center cannot revoke under current governance rules. The framework distinguishes \emph{capacity scaling within a topology} from \emph{changing the topology itself}.

A useful diagnostic is to ask what kind of constraint is binding. If delays arise from test execution, review bandwidth, or architecture comprehension, additional tools or HACs may be sufficient. If delays arise because one anchor is being asked to accept risks that require durable independent judgment---for example regulatory separation of duties, product-domain ownership, safety assurance, or contractual governance---then capacity expansion alone does not remove the organizational constraint. The problem has moved from execution scaling to responsibility design.

When
\[
\text{Generation Rate}>\text{Verification Rate},
\]
a Verification Queue forms. As the queue grows, the responsible subject has stronger incentives to rely on summaries, automated reviewers, and rapid approvals, which may weaken genuine engineering judgment. The Verification Queue is an observable backpressure signal in agentic software production.

\paragraph{Evidence independence.}
A single responsibility center lacks some of the professional heterogeneity that arises naturally in larger teams. If coding, testing, and reviewing agents share a base model, specification, and context, multiple agents may miss the same defect.

Single-center SE can increase Evidence Independence through deterministic static analysis, property-based testing, mutation testing, hidden tests, heterogeneous model families, real-user validation, and external professional review. External expert networks matter not only as extra execution resources but as sources of independent judgment.

\subsection{Human Attention, Architecture, and Product Judgment}
As agents expand execution, the responsible person's attention becomes an explicit system bottleneck. Major attention costs include
\[
\text{Intent}+\text{Specification}+\text{Review}+\text{Integration}+\text{Incident}+\text{Context Recovery}.
\]
Change Locality directly influences scalability. If a small feature forces the final anchor to reconstruct many modules, protocols, data models, and agent traces, high automation does not translate into stable leverage.

A scalable single-center architecture depends on keeping the semantic scope, review scope, failure radius, and long-term responsibility obligations of a typical TC bounded.

\paragraph{Product judgment and complete cost.}
As construction cost falls, the decision to create a change becomes more important. The responsible subject must ask whether the problem being solved is sufficiently evidenced, whether user value justifies migration and compatibility costs, whether the change creates new support or security obligations, and whether its long-term responsibility obligations are affordable.

A more complete cost model for a TC is
\[
\text{Creation}+\text{Verification}+\text{Integration}+\text{Operation}+\text{Long-Term Responsibility}.
\]
Reducing Creation does not proportionally reduce lifecycle cost. Resource optimization should therefore focus on
\[
C_{\mathrm{Accepted\ TC}}
\]
rather than on the price of an individual model call.

\subsection{Operations and Continuity}
Known, low-risk, reversible operational actions can be pre-authorized to agents: collecting logs, estimating impact, diagnosing known failures, or preparing rollback options. High-impact, uncertain, or irreversible actions escalate to $R^*$.

If old SDKs begin failing on the new \texttt{expired} state after release, an agent may automatically quantify affected versions, generate compatibility patches, and prepare rollback plans. The choice among rollback, forward fix, or temporary semantic mapping remains a risk-acceptance decision. The objective is an explicit boundary between pre-authorized automation and high-impact human acceptance.

\paragraph{Takeoverability and anchor-absence resilience.}
Because the final responsibility center is a natural single point, two distinct continuity properties matter.

\textbf{Takeoverability} measures whether, during formal responsibility transfer, a new authorized subject can identify the current baseline, acquire necessary permissions, stop high-risk agents, restore service, and assume required responsibilities.

\textbf{Anchor-Absence Resilience} measures whether the system can continue safely, degrade automatically, or wait in a controlled state while the primary anchor is temporarily unavailable, without immediately requiring formal responsibility transfer.

The first concerns responsibility transfer; the second concerns temporary absence. High automation creates neither property by itself. Both depend on externalizing critical knowledge, state, authority, and emergency procedures.

\subsection{When Single-Center Is No Longer the Right Topology}
A single responsibility center can become mismatched with the risk structure as the software develops independent high-risk domains, institutional separation-of-duty requirements, continuous on-call needs, or regulatory obligations.

If new governance rules create multiple mutually independent Responsibility Anchors, then
\[
\text{Single-center SE}\rightarrow\text{Multi-anchor Team SE}.
\]
This is a Responsibility Topology change.

Hard governance triggers---such as regulation, separation of duties, or independent safety/security acceptance---determine the topology directly. Without such a trigger, the issue is economic and organizational: a new anchor becomes attractive only when the costs and continuity risks of one center justify an independently responsible domain. Section~\ref{sec:metrics} develops this distinction and the transition condition.

\paragraph{Diagnostic criteria.}
Table~\ref{tab:single-center-failures} summarizes the major single-center failure mechanisms.
\begin{table}[H]
\centering
\small
\caption{Major failure mechanisms in single-center SE.}
\label{tab:single-center-failures}
\begin{tabularx}{\linewidth}{@{}p{0.27\linewidth}p{0.32\linewidth}X@{}}
\toprule
\textbf{Failure mechanism} & \textbf{Manifestation} & \textbf{Root cause} \\
\midrule
Verification Queue & Candidates continuously accumulate & Generation exceeds verification capacity \\
Same-origin assurance & Multiple agents share the same blind spot & Insufficient Evidence Independence \\
Cognitive overload & The anchor cannot understand material changes & Change scope exceeds cognitive budget \\
Lifecycle obligations & Current changes are cheap but long-term obligations accumulate & Lifecycle cost is underestimated \\
Anchor dependency & System capability collapses when the anchor leaves & Insufficient Takeoverability / Absence Resilience \\
\bottomrule
\end{tabularx}
\end{table}

The effective scale of single-center SE is ultimately determined by how many Accepted TCs one responsibility center can continuously understand, verify, integrate, finance, and sustain through their lifecycles.

\section{Multi-Anchor Team Software Engineering}
\label{sec:team}

Section~\ref{sec:methods} defined the common engineering methods. This section studies the practically important case in which distributed HAC execution coexists with multiple independent Responsibility Anchors. The coordination and authority pressures should not be conflated: context divergence is primarily induced by distributed execution, whereas joint acceptance and Responsibility Closure follow from distributed authority. Multi-anchor governance makes explicit authority, versioning, and invalidation semantics necessary when cross-anchor closure depends on shared engineering facts. The running order-state example connects Authoritative State, Context Invalidation, delegation, evidence, and responsibility closure.

\subsection{Authoritative State and Context Coherence}
TSP already demonstrates that the quality of individual software processes does not add up automatically to team-process quality~\cite{ref5}. In Agent SE, a team is composed of
\[
HAC_1+HAC_2+\cdots+HAC_n,
\]
where each HAC has independent agents, memory, tools, permissions, and budgets. Distributed HAC execution creates local-memory and evidence-coordination pressures; when the organization also contains multiple independent anchors, these coexist with a responsibility-closure problem. In the combined case analyzed here, the team faces three questions:
\begin{enumerate}
\item how Local Memory relates to Global Engineering Truth;
\item how the Delegation Graph aligns with the Responsibility Graph; and
\item how local evidence becomes a trustworthy system-level basis for acceptance.
\end{enumerate}
These are central research concerns for future human--agent team software-process research.

\paragraph{Shared Authoritative Engineering State.}
An authoritative engineering state is useful under either topology. Multiple independent anchors make explicit authority, versioning, and invalidation semantics structurally necessary for cross-anchor closure: the team need not share all local memory, but it does need a jointly recognized set of formal engineering facts. We call this the \textbf{Shared Authoritative Engineering State}.

Typical content includes Product Intent, Current Specification, Architecture Decisions, API/Data Contracts, Global Invariants, Ownership, Accepted TCs, Risk State, and Release State.

An Authoritative-State item can be represented as
\[
ASE=\left\langle\text{Content},\text{Proposer},\text{Acceptance Anchors},\text{Version},\text{Effective Time},\text{Invalidation Policy}\right\rangle.
\]
The proposer may be a HAC or an agent. A proposal gains organizational authority only after relevant anchors accept it and it is committed in versioned form:
\[
\text{Proposal}\rightarrow\text{Authorized Acceptance}\rightarrow\text{Versioned Commit}.
\]
The mechanism distinguishes information that is shareable from engineering facts that are actually binding.

The stronger requirement follows from the multi-anchor topology. When several independent anchors govern different claims about the same system, no participant's local memory can serve as the organization's implicit source of truth. A local note may be accurate, widely retrieved, and repeatedly copied while still lacking the authority to constrain another HAC. Conversely, once a semantic decision is accepted into the authoritative state, dependent work must be able to determine which version was effective and whether its own assumptions remain valid. Authority and retrievability are distinct dimensions.

This creates a second-order problem that is much less visible in single-center work: accepted information changes the validity of work already in flight. Suppose the order-state contract advances from v3 to v4. A backend HAC already working against v4 may continue; a client HAC using v3 semantics must refresh and replan; a payment HAC whose late-callback evidence depended on the old contract must reverify; and an analytics task that would require unsupported historical inference should stop. The engineering consequence of authoritative-state change is thus not merely synchronization of documents but \emph{selective invalidation of distributed reasoning and evidence}.

\paragraph{Context coherence.}
Context Coherence asks whether a formal engineering action can identify the Authoritative Facts on which it depends, their versions, and whether those facts remain valid.

Different HACs may continue to have different contexts:
\[
\mathrm{Context}_A\neq\mathrm{Context}_B.
\]
Coherence does not require all agents to hold identical context. Team state can instead be classified by consistency requirements.

\paragraph{Strong consistency.}
Strong consistency is appropriate for Global Invariants, public API contracts, formal ownership, production schemas, and release state.

\paragraph{Eventual consistency.}
Eventual consistency is appropriate for exploratory plans, preliminary analyses, and non-critical state that can tolerate asynchronous synchronization.

\paragraph{Local state.}
Local state includes personal-agent scratch memory, provisional hypotheses, and unaccepted alternatives.

Accordingly,
\[
\text{Shared Memory}\neq\text{Shared Authoritative State}.
\]

Context Coherence is achieved when a formal action can identify the authoritative facts on which it depends, the versions of those facts, and the risk-appropriate response when they change. This definition deliberately avoids requiring globally strong consistency for every piece of information. Exploratory plans and temporary analyses can remain local or eventually consistent; public interfaces, accepted invariants, ownership, and release state may require stronger coordination. The framework therefore treats consistency as a property chosen by the risk and authority semantics of the information, not as a universal storage policy.

\subsection{Context Invalidation and End-to-End Closure}
When authoritative engineering state changes from
\[
S_v
\]
to
\[
S_{v+1},
\]
in-flight tasks depending on the earlier version should select one of the following actions according to impact:
\[
\operatorname{Action}(Task_i)\in\{\text{Continue},\text{MarkStale},\text{Refresh},\text{Replan},\text{Reverify},\text{Stop}\}.
\]

Continue means the authoritative change is immaterial to the task. MarkStale permits further exploration but forbids promotion to Eligible. Refresh updates context only. Replan changes the execution plan. Reverify reconstructs evidence. Stop terminates a task whose effective objective is no longer valid.

\begin{figure}[H]
\centering
\includegraphics[width=0.96\linewidth]{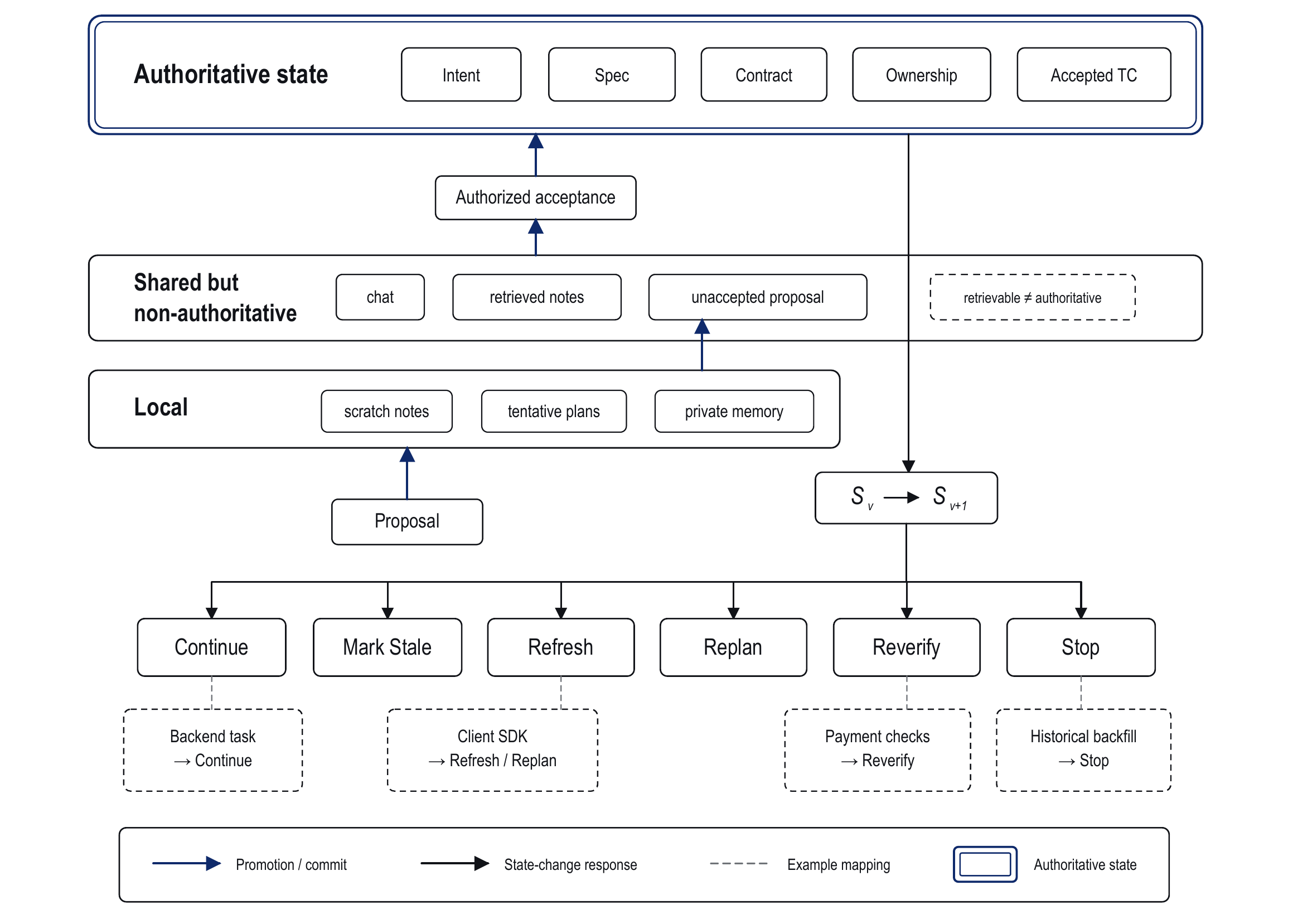}
\caption{From local information to Authoritative Engineering State and then to Context Invalidation. Shareable or retrievable information becomes binding only through authorized acceptance and versioned commitment. Context invalidation can arise under either topology; multi-anchor governance makes the authority/version path explicit for cross-anchor closure.}
\label{fig:context-invalidation}
\end{figure}

For any affected TC, this policy operationalizes
\[
\mathrm{Eligible}(x,S_v)\not\Rightarrow\mathrm{Eligible}(x,S_{v+1}):
\]
eligibility must be revalidated after the relevant authoritative facts change. Figure~\ref{fig:context-invalidation} summarizes the corresponding promotion and state-change-response paths.

After the order-status contract changes from v3 to v4, a backend task already based on v4 can Continue. A client task still interpreting every closed order as \texttt{cancelled} must Refresh and Replan. Payment race analysis must Reverify. A historical automated backfill task invalidated by the new semantic decision can Stop.

\paragraph{Complete team closure.}
The Delegation Graph describes how work is decomposed, assigned, and executed. The Responsibility Graph describes which subjects possess formal acceptance authority for which responsibility domains. The two can differ substantially:
\[
\text{Delegation Graph}\neq\text{Responsibility Graph}.
\]
A characteristic system-level failure is
\[
\text{Execution Closure}\neq\text{Responsibility Closure}.
\]

\begin{figure}[H]
\centering
\includegraphics[width=0.99\linewidth]{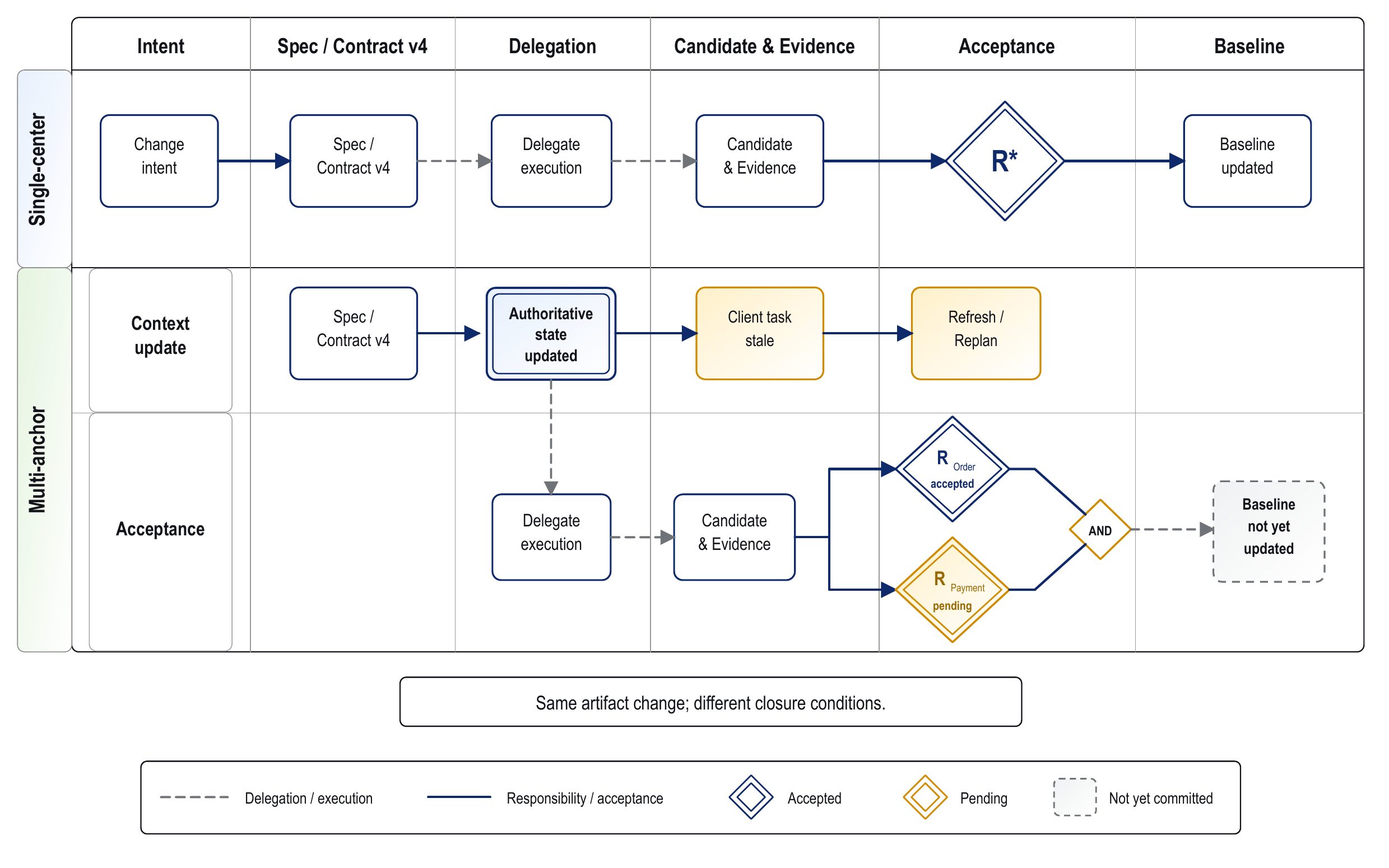}
\caption{The same order-state change has different closure conditions under single-center and multi-anchor topologies. The multi-anchor path combines a context-invalidation example caused by distributed execution with waiting for independent acceptance; only the latter is topology-specific.}
\label{fig:topology-swimlane}
\end{figure}

\paragraph{End-to-end closure of the order-state change.}
Figure~\ref{fig:topology-swimlane} contrasts the closure paths for the same artifact change under single-center and multi-anchor governance; the following sequence then walks the multi-anchor case through Authoritative State, invalidation, delegation, evidence, and acceptance.

\textbf{Step 1 -- Proposal.} A Product HAC proposes splitting \texttt{cancelled} into \texttt{cancelled} and \texttt{expired}, producing a new Order Status Contract proposal.

\textbf{Step 2 -- Authoritative State.} The Order and Payment Responsibility Anchors separately accept the domain semantics and payment-timeout semantics. After authorized acceptance, the proposal becomes $OrderStatusContract=v4$ and enters Shared Authoritative Engineering State.

\textbf{Step 3 -- Context Invalidation.} The Client HAC still uses v3 semantics and performs $Refresh+Replan$. A prior late-callback analysis in the Payment HAC performs Reverify. A historical automated backfill task contradicted by the new decision performs Stop.

\textbf{Step 4 -- Delegation.} Backend, Client, and Payment HACs receive the common Task \& Resource Contracts defined in Section~\ref{sec:methods}. The topology-specific point is not a new contract schema: delegation may be orchestrated centrally while authority to accept the resulting change remains distributed across independent anchors.

\textbf{Step 5 -- Candidate.} Each HAC outputs a Candidate Change and Local Evidence.

\textbf{Step 6 -- R4 failure example.} Suppose the Payment Candidate implements \texttt{expired}, but no evidence covers a successful payment callback arriving after automatic timeout closure. The candidate triggers R4. It cannot become Eligible even if the code compiles and ordinary unit and SDK tests all pass.

\textbf{Step 7 -- R6 failure versus Acceptance Pending.} If no valid Responsibility Anchor exists for the Payment domain under current governance rules, R6 fails. If $R_{\mathrm{Payment}}$ validly exists but has not yet formally accepted the change, R6 does not fail. The change may be in
\[
\mathrm{Eligible}=\mathrm{true},\qquad \mathrm{Accept}_K=\mathrm{false},
\]
that is, Eligible / Acceptance Pending.

\textbf{Step 8 -- Multi-anchor acceptance.} Once the evidence is complete and eligibility passes,
\[
K=\{R_{\mathrm{Order}},R_{\mathrm{Payment}}\},
\]
\[
\mathrm{Accept}_K(x)=\mathrm{Accept}_{\mathrm{Order}}(x)\land\mathrm{Accept}_{\mathrm{Payment}}(x),
\]
\[
\mathrm{AcceptedTC}(x)\iff\mathrm{Eligible}(x)\land\mathrm{Accept}_K(x).
\]

\begin{figure}[H]
\centering
\includegraphics[width=0.96\linewidth]{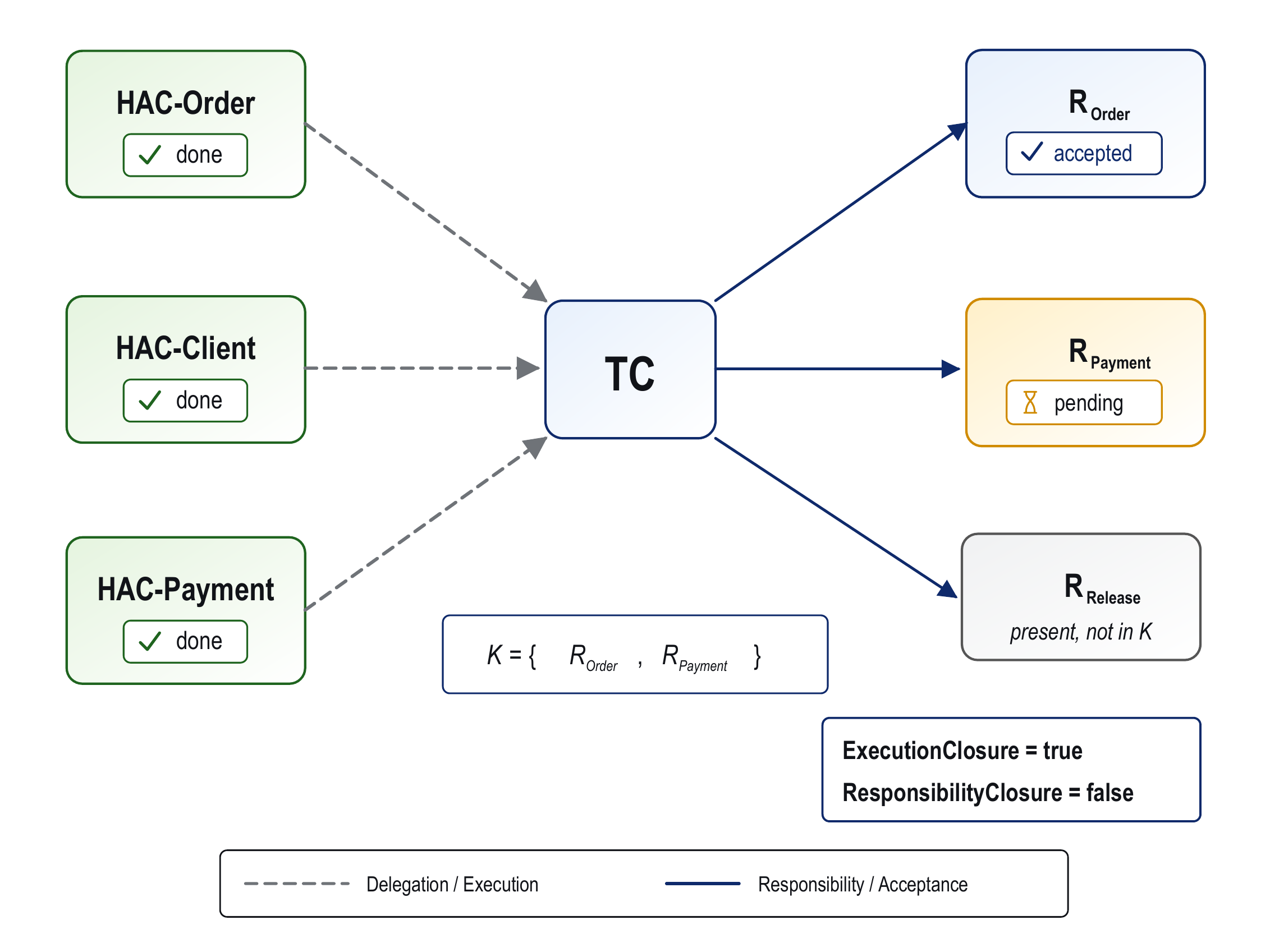}
\caption{Execution Closure does not imply Responsibility Closure. All three HACs have completed execution, but the Payment Anchor has not accepted, so the TC cannot enter the baseline.}
\label{fig:dual-graphs}
\end{figure}

\textbf{Step 9 -- Baseline and operation.} The Accepted TC enters the authoritative baseline. Legacy-SDK behavior, late callbacks, refund processing, and analytics continue to generate Runtime Evidence. If runtime facts invalidate the assumptions on which acceptance depended, a new Intent and TC are created. Figure~\ref{fig:dual-graphs} makes the separation between execution and responsibility closure explicit for the running example.

\subsection{Delegation, Responsibility, and Layered Evidence}
Teams can exploit different HAC contexts, expertise, and responsibility positions to increase evidence independence. A risk-adaptive structure can be organized as
\[
\text{Local}\rightarrow\text{Cross-cell}\rightarrow\text{Team}\rightarrow\text{Organization/External}.
\]
Local verification is cheapest but most correlated. Cross-cell verification asks a different HAC to inspect the change. Team-level verification addresses architecture, contracts, and global invariants. Organization/external verification addresses high-risk areas such as security, compliance, and finance.

R4 remains Evidence Adequacy; Responsibility Topology does not alter its definition. What changes is the provenance, independence, and relevance required of the evidence used to satisfy R4. For higher-risk cross-domain changes, cross-cell or external evidence can reduce the correlation of same-origin errors and shared blind spots, provided that the evidence remains relevant to the changed claims and acceptance scope.

Agent scope can similarly be described as Personal, Team, and Organization scope. These levels describe execution scope; they do not determine formal responsibility structure.

\subsection{Team Throughput, Organizational Memory, and the Augmented Conway Hypothesis}
With multiple humans and multiple agents, effective concurrency is constrained by
\[
C_{\mathrm{multi}}\le\min\left(C_D,B_{\mathrm{local}},B_{\mathrm{cross}},B_{\mathrm{integration}},B_{\mathrm{architecture}},B_{\mathrm{economic}}\right).
\]
Here $C_D$ is the decomposability-induced throughput ceiling; under a fixed workload, all terms are expressed in a common throughput unit such as admissible TCs per unit time. Higher local productivity in individual HACs can still coexist with lower system-level Trustworthy Change throughput.

Team project state should also distinguish
\[
\begin{aligned}
\text{Planned}&\rightarrow\text{Delegated}\rightarrow\text{Executing}\rightarrow\text{Candidate}\\
&\rightarrow\text{Eligible}\rightarrow\text{Acceptance Pending}\rightarrow\text{Accepted}.
\end{aligned}
\]
Organizational memory should preferentially preserve current decisions, rationales, specifications, invariants, ownership, Accepted TCs, and known risks. Chat transcripts and agent traces can be retained for audit but do not automatically acquire authoritative status.

\paragraph{Operationalizing the augmented Conway hypothesis.}
Conway's classic observation connects organizational communication structures to system structures~\cite{ref2}. We propose a testable extension: beyond human communication, Agent Delegation and Context Dependency may also influence software co-change and integration structure.

Define
\[
G_H=\text{Human Communication Graph},\qquad G_D=\text{Agent Delegation Graph},
\]
\[
G_C=\text{Context Dependency Graph},\qquad G_S=\text{Software Dependency / Co-change Graph}.
\]
Because these graphs may have different node types, empirical work should project them onto a shared Task, Module, or TC space rather than adding the graphs directly.

The empirical question is whether, when two engineering units must coordinate because of software dependency, Human Communication, Agent Delegation, and Context Sharing provide corresponding coordination paths. Candidate outcomes include Rework Rate, Integration Conflict Rate, Cross-HAC Replan Rate, Context Refresh/Revalidation Rate, Stale-Context Escape Rate, TC Lead Time, and Defect/Rollback Rate.

We call the possibility that Agent Delegation and Context Dependency add explanatory power after controlling for Human Communication the \textbf{Augmented Conway Effect hypothesis}.

\subsection{Failure Modes and a Human--Agent Team Process Research Direction}
Table~\ref{tab:team-failures} summarizes the principal failure modes in the combined distributed-HAC, multi-anchor setting.
\begin{table}[H]
\centering
\small
\caption{Major failure modes when distributed HAC execution coexists with multi-anchor governance.}
\label{tab:team-failures}
\begin{tabularx}{\linewidth}{@{}>{\raggedright\arraybackslash}p{0.34\linewidth}X@{}}
\toprule
\textbf{Failure mechanism} & \textbf{Typical manifestation} \\
\midrule
Context Staleness & Agents act on outdated specifications \\
Unauthorized Truth & An unaccepted proposal is treated as authoritative fact \\
Delegation--Responsibility\newline Misalignment & Execution is complete but responsibility is not closed \\
Cross-cell Semantic Conflict & Artifacts merge while semantics conflict \\
False Independence & Multiple reviewers share the same model and blind spots \\
Verification Congestion & Local candidates exceed cross-HAC verification capacity \\
Invalidation Storm & Frequent authoritative changes force repeated task rework \\
Responsibility Orphan & An anchor exits and a responsibility domain loses a valid acceptor \\
\bottomrule
\end{tabularx}
\end{table}

In summary, multi-anchor Team SE combines common Agent SE methods with distributed-execution coordination and topology-specific responsibility closure. The mechanisms above motivate a future research direction on human--agent team software processes. Within the scope of this paper, Authoritative State, Context Coherence, the dual-graph relation, and cross-cell evidence are recurring concerns; we do not prescribe a fixed team process script or introduce another process standard.

\section{A Reference Tool Architecture for Agent Software Engineering}
\label{sec:tooling}

\paragraph{Three layers and their theoretical roles.}
Section~\ref{sec:methods} defined common methods such as Task Contracts, evidence, and sandboxing, while Section~\ref{sec:team} defined Authoritative State, Context Coherence, and Responsibility Graphs. This section does not propose a component-by-component product design. It shows how those theoretical objects can be projected onto existing software-engineering infrastructure.

\begin{figure}[H]
\centering
\includegraphics[width=0.94\linewidth]{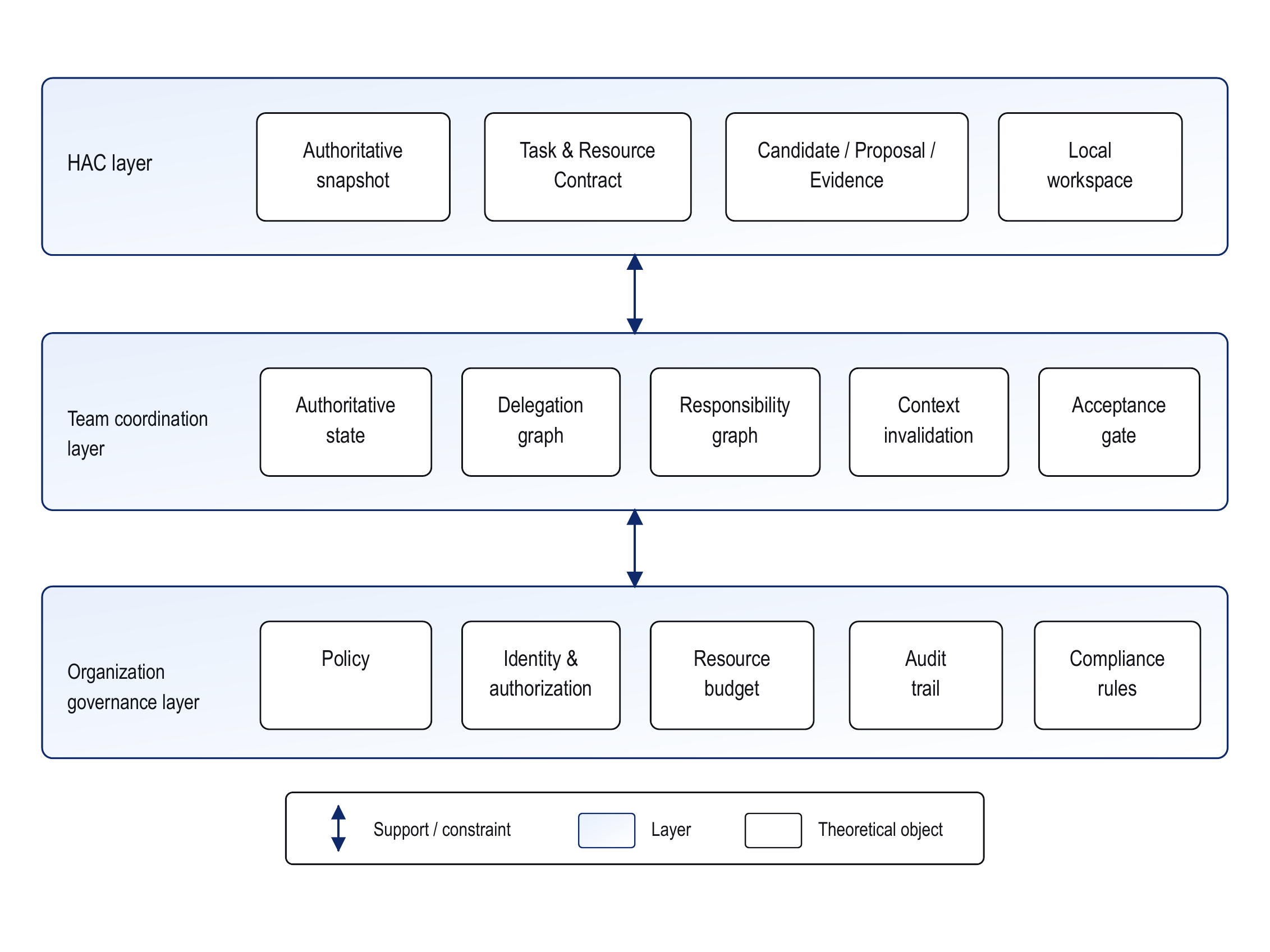}
\caption{Three-layer reference architecture for Agent SE. The figure shows where theoretical objects can be represented, not a mandatory product decomposition.}
\label{fig:tool-architecture}
\end{figure}

Tool capabilities can be organized by scope:
\[
\text{HAC/Personal}\rightarrow\text{Team}\rightarrow\text{Organization}.
\]
Figure~\ref{fig:tool-architecture} projects these scopes onto a three-layer reference architecture.

The HAC layer primarily supports local execution, permission boundaries, and candidate evidence. The Team layer supports facts shared across HACs, Context Invalidation, responsibility acceptance, and integration. The Organization layer supports identity, policy, compliance, model/tool governance, and cost control. Concrete platforms may merge or split these capabilities; the framework does not require one service per theoretical object.

\paragraph{Mapping industrial primitives to engineering objects.}
Current agent-development platforms already provide repository-instruction files such as \texttt{AGENTS.md} and \texttt{CLAUDE.md}, skills, hooks, MCP tools, sandboxes, subagents, and specification workflows. A September 2026 source-code study of eleven production coding-agent harnesses finds recurring runtime subsystems for execution loops, tool use, context management, safety controls, orchestration, and extension surfaces~\cite{ref46}. This makes the status of the tooling layer important to state explicitly: these primitives are not atomically novel contributions of this paper. They are the existing substrate on which TC, Responsibility Topology, Authoritative State, and Context Invalidation can be operationalized over a complete software change. Table~\ref{tab:primitive-mapping} maps representative industrial primitives to framework roles.

\begin{table}[H]
\centering
\small
\caption{Mapping industrial agent primitives to the framework.}
\label{tab:primitive-mapping}
\begin{tabularx}{\linewidth}{@{}p{0.32\linewidth}X@{}}
\toprule
\textbf{Industrial primitive} & \textbf{Primary role in the framework} \\
\midrule
AGENTS.md / CLAUDE.md / Repository Rules & Part of an Authoritative State Snapshot or task-local context; authority depends on acceptance and versioning \\
Spec / Plan & Intent \& Specification \\
Skills / Custom Agents & Delegated execution capability \\
MCP / Tools & Tool capability in a Task Contract \\
Hooks / Policy Hooks & Delegation guard, Stop Condition, deterministic verification \\
Sandbox / Worktree / Cloud VM & Permission boundary and Candidate isolation \\
Subagents / Agent Team & Execution nodes in a Delegation Graph \\
Tests / CI / merge-readiness package & Evidence and eligibility inputs \\
CODEOWNERS / approval policy / IAM & One form of technical evidence for a Responsibility Graph \\
Usage / token / model routing & Resource Envelope and bounded-capacity control \\
\bottomrule
\end{tabularx}
\end{table}

\paragraph{Implementation plausibility.}
The reference architecture primarily adds state semantics and governance relations to infrastructure that already exists. It does not depend on unavailable foundational computing capabilities. Table~\ref{tab:implementation-plausibility} summarizes how the core mechanisms can reuse existing infrastructure.

\begin{table}[H]
\centering
\small
\caption{Implementation plausibility of the core mechanisms.}
\label{tab:implementation-plausibility}
\begin{tabularx}{\linewidth}{@{}p{0.24\linewidth}p{0.31\linewidth}X@{}}
\toprule
\textbf{Mechanism} & \textbf{Reusable infrastructure} & \textbf{Additional Agent-SE semantics} \\
\midrule
Authoritative State & Git, artifact stores, CMDB & Acceptance, version, effective time, invalidation \\
Context Coherence & Dependency graphs, event buses & Task--state dependency and staleness \\
Responsibility Graph & IAM, CODEOWNERS, approval matrices & Residual-risk acceptance authority \\
Task Contract & Sandboxes, policy engines, workflows & Permission, budget, stop, handoff \\
Evidence Graph & CI, test reports, static analysis & Claim--evidence relations and Evidence Independence \\
TC Pipeline & CI/CD, change management & Candidate $\rightarrow$ Eligible $\rightarrow$ Accepted TC state machine \\
\bottomrule
\end{tabularx}
\end{table}

This mapping establishes \emph{implementation plausibility}: prototypes can be built by extending existing engineering infrastructure. It is not empirical validation of utility, cost, organizational impact, or developer experience.

\section{Measurement, Engineering Economics, and Project Control}
\label{sec:metrics}

\subsection{Measurement Model and Accepted-TC Accounting}
This section translates the structural constraints in Sections~\ref{sec:single-center} and~\ref{sec:team} into observable engineering variables. Verification queues, human attention, and anchor dependency in single-center SE must be measurable; context invalidation and cross-cell verification in distributed HAC execution, and Responsibility Closure in multi-anchor governance, must likewise become project-state and economic variables.

The resource vectors, flow measures, and cost variables defined here have two purposes: to provide operational quantitative tools for single-center and multi-anchor SE, and to establish candidate measurement models for the empirical tests in Section~\ref{sec:agenda}. This follows the long-standing goal-driven measurement principle in software engineering. GQM derives Questions from explicit Goals and then derives Metrics from those Questions rather than beginning with whatever activity data happen to be easy to collect~\cite{ref37}.

The section operationalizes the basic claim of Bounded-Capacity Agent SE: digital execution can be elastically expanded, but verification, integration, Human Attention, and responsibility acceptance remain capacity-constrained.

The central measurement principle is to derive observables from the constructs rather than to select convenient activity counters first. For example, Responsibility Topology can be operationalized through the number and independence of required acceptance anchors for a TC, then related to outcomes such as acceptance latency, cross-domain waiting time, or unresolved long-term responsibility obligations. Context Invalidation can be operationalized through the number of affected in-flight tasks, the actions issued to those tasks, and the rate at which stale context escapes into later verification or integration. Evidence independence can be represented through diversity of method, context, criterion, and responsibility source rather than by simply counting reviewer agents.

These mappings also define what should \emph{not} be interpreted as evidence for the theory. A high commit count does not imply high TC throughput; many agent trajectories do not imply many independent verification sources; and a large team does not imply a multi-anchor topology. Accepted TC is used as the accounting boundary precisely because it is closer to the theoretical object than raw activity measures. The measurement layer is therefore part of the operationalization of the theory, not a separate productivity framework.

\paragraph{Accepted-TC accounting.}
Agents lower the cost of producing code, tests, and documentation, further weakening the relationship between LOC, commit count, pull requests, agent-task count, and actual software value.

We use \textbf{Accepted TC} as the accounting boundary for trustworthy output. Accepted-TC accounting is intended for within-project lifecycle analysis; it is not a cross-project scalar of software value. Its primary purpose is to distinguish changes that have entered the formal software baseline from work that remains Candidate, Rejected, Abandoned, or Eligible-but-not-Accepted.

The SPACE framework emphasizes that developer productivity is multidimensional and cannot be represented by a single activity or output metric~\cite{ref6}. Accepted-TC Accounting retains this caution: TC is an accounting boundary, not a universal productivity score.

\subsection{Resources, Cost, and Agentic Effort}
Using the resource definition from Section~\ref{sec:methods},
\[
R=\left\langle\text{Token},\text{Compute},\text{Tool},\text{Wall-Clock Time},\text{Human Attention}\right\rangle.
\]
Tokens capture model inference and context resources; Compute includes builds, tests, simulations, and agent runtime; Tool includes APIs, databases, and external services; Wall-Clock Time is elapsed delivery time; Human Attention includes clarification, review, integration, exception handling, and responsibility acceptance.

\[
\text{Token}\neq\text{Human Hours}.
\]
Tokens are better interpreted as a digital execution resource.

Candidate project-level measures include
\[
\operatorname{TokenPerAcceptedTC}=\frac{\text{Total tokens}}{\#\{\text{Accepted TC}\}},
\]
\[
\operatorname{HumanAttentionPerAcceptedTC}=\frac{\text{Total human attention}}{\#\{\text{Accepted TC}\}},
\]
and
\[
C_{\text{rejected changes}}=\operatorname{Cost}(\text{candidates not accepted}).
\]
These are most useful for longitudinal comparison within a project, experiments on agent configurations, and analysis of architecture evolution.

\paragraph{Architecture, context, and verification cost.}
Change Locality connects architecture to agent-engineering cost. Larger Artifact Span is expected to increase integration scope; larger Context Span may increase token and review cost; larger Responsibility Span may increase cross-cell and acceptance cost.

Teams should additionally observe Context Staleness and Revalidation Load. For example,
\[
\operatorname{ContextStalenessRate}=
\frac{\#\{\text{active tasks affected by stale authoritative context}\}}
{\#\{\text{active tasks}\}}.
\]

Evidence Independence also creates a cost--risk trade-off. External experts, different-model review, cross-cell verification, and formal verification are usually more expensive than local self-check. Verification investment should therefore scale with risk and impact rather than being maximized uniformly.

\paragraph{Agentic effort estimation.}
Agent-era project control still requires estimates of how many resources a change will consume. Candidate predictors include Ambiguity, Context Size, Change Locality, Tool Steps, Risk, Verification Requirement, and Responsibility Topology.

We hypothesize that agentic effort is jointly influenced by specification ambiguity, required context, change locality, tool steps, risk, verification requirements, and responsibility topology. No particular functional form, monotonicity, or unit scale is assumed. Section~\ref{sec:agenda} treats the relation as an empirical estimation problem.

Ambiguity also lacks a single reliable measure. Candidate proxies include Clarification Count, Unresolved Specification Slots, Mutually Exclusive Interpretations, Candidate Semantic Divergence, and Replan Rate induced by semantic clarification. Such proxies should be calibrated by model and harness.

\subsection{Concurrency, Flow Control, and Topology Economics}
Single-center capacity is constrained by
\[
C_{\mathrm{single}}\le\min\left(C_D,B_{\mathrm{integration}},B_{\mathrm{verification}},B_{\mathrm{economic}}\right),
\]
while multi-anchor team capacity is constrained by
\[
C_{\mathrm{multi}}\le\min\left(C_D,B_{\mathrm{local}},B_{\mathrm{cross}},B_{\mathrm{integration}},B_{\mathrm{architecture}},B_{\mathrm{economic}}\right).
\]
Here $C_D$ is the decomposability-induced throughput ceiling. Under a fixed workload, every term in each bound denotes an effective throughput ceiling expressed in a common unit such as admissible TCs per unit time. These expressions are schematic rather than calibrated performance laws. Agent Execution Concurrency remains bounded by downstream verification, integration, architecture, and economic capacity.

\begin{figure}[H]
\centering
\includegraphics[width=0.99\linewidth]{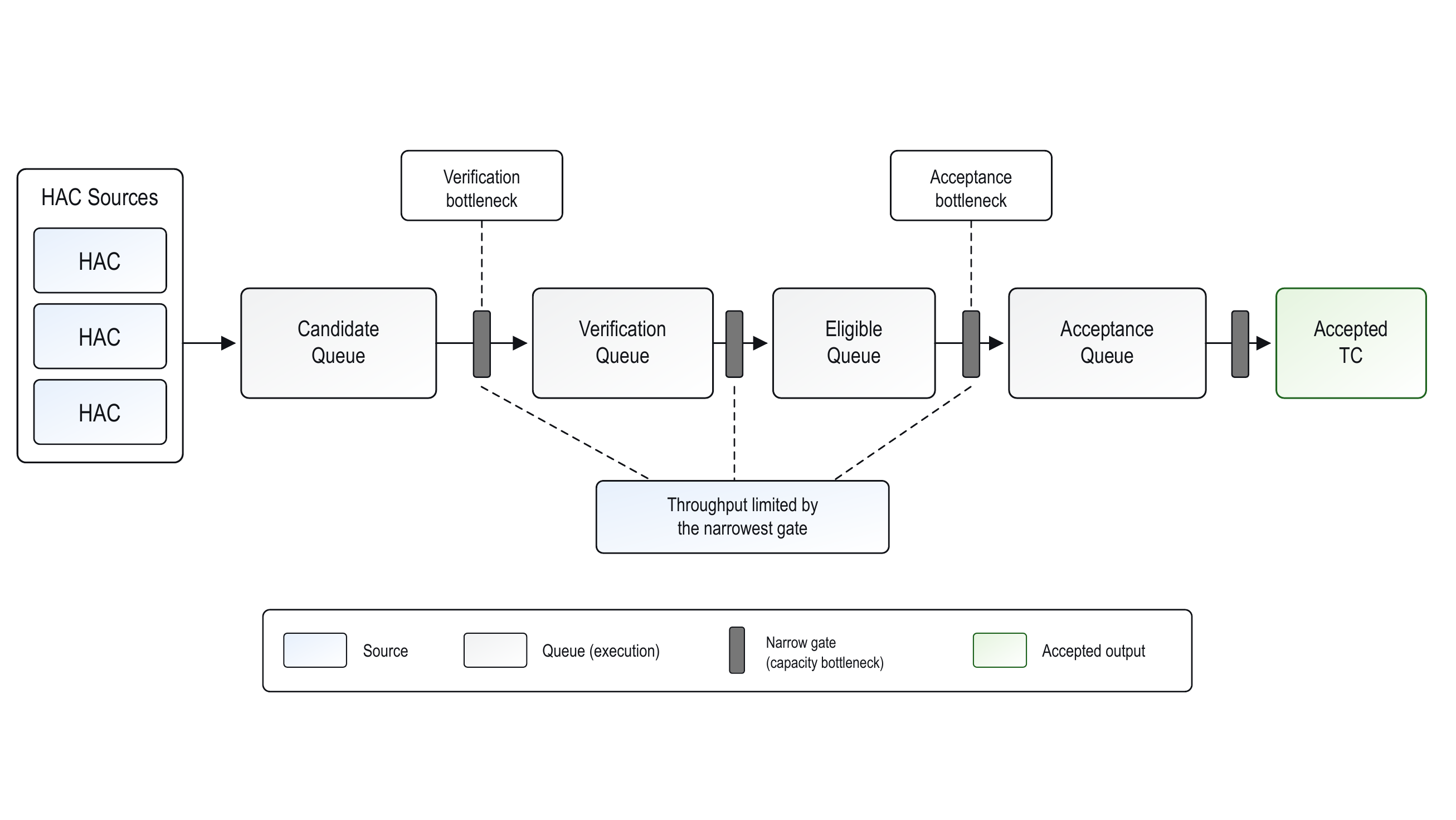}
\caption{Trustworthy Change flow and potential bottlenecks. As agent generation throughput increases, verification and responsibility acceptance can become new queues.}
\label{fig:tc-flow-queue}
\end{figure}

Project dashboards should track the Trustworthy Change Flow; Figure~\ref{fig:tc-flow-queue} illustrates the verification and acceptance queues that may become binding.

METR's 2026 revision of developer-productivity experiments similarly notes that multi-agent concurrency, task selection, and agent usage patterns make simple wall-clock comparisons harder to interpret~\cite{ref28}.

\paragraph{Economic pressure on responsibility topology.}
Responsibility Topology is defined first by governance and residual-risk acceptance structure, not by direct cost minimization. A move from single-center to multi-anchor Team SE must therefore distinguish two types of condition.

\paragraph{Hard governance triggers.}
Law or regulation, professional qualifications, separation of duties, independent security or safety acceptance, and contractual governance requirements can directly mandate independent responsibility structures. In these cases Team Topology is a governance requirement; economic efficiency is secondary.

\paragraph{Economic transition pressure.}
Absent a hard governance trigger, the costs of maintaining a single responsibility center can be compared with those of a multi-center structure:
\[
C_{\mathrm{single}}=C_{\mathrm{verification\ queue}}+C_{\mathrm{anchor\ attention}}+C_{\mathrm{external\ assurance}}+C_{\mathrm{continuity\ risk}}+C_{\mathrm{delay}}+\cdots,
\]
\[
C_{\mathrm{multi}}=C_{\mathrm{cross-cell}}+C_{\mathrm{state\ synchronization}}+C_{\mathrm{multi-anchor\ acceptance}}+C_{\mathrm{governance}}+\cdots.
\]
The scalar comparison assumes that heterogeneous costs have been mapped by a project-specific valuation model into a common economic unit; otherwise they should remain a cost vector.

Define
\[
\Delta C_{\mathrm{topology}}=C_{\mathrm{single}}-C_{\mathrm{multi}}.
\]
If $\Delta C_{\mathrm{topology}}>0$, a team structure \emph{may} have an economic advantage. The inequality does not automatically justify creation of a new Responsibility Anchor.

If the Verification Queue reflects only a shortage of mechanical review capacity, single-center SE can add agents, external reviewers, or delegated HACs while preserving one final responsibility center. A new anchor requires an additional condition: a responsibility domain with a legitimate, stable acceptance right that is not normally unilaterally overridable by the current anchor.

\subsection{Human Attention and Primary Metrics}
Agents may reduce implementation hours while increasing review fragmentation, context switching, and approval frequency. Human Attention should therefore be observed structurally---for example Context Switch Count, Deep-Work Interval, and Time-to-Regain-Context---rather than only as total labor hours. Human-attention measurement should prefer low-intrusion workflow proxies, aggregated telemetry, or voluntary self-report where possible; empirical studies should treat privacy, surveillance effects, and measurement overhead as part of the measurement design.

Every Accepted TC also creates long-term responsibility obligations: compatibility, maintenance, security updates, data obligations, customer support, incidents, and eventual exit. Classical software engineering economics already places lifecycle cost, maintenance, risk, and project decisions in one analytical frame~\cite{ref38}; Agent SE adds tokens, agent runtime, verification queues, and human-attention costs.

\[
C_{\mathrm{lifecycle}}(TC)=C_{\mathrm{creation}}+C_{\mathrm{verification}}+C_{\mathrm{integration}}+C_{\mathrm{operation}}+C_{\mathrm{long\text{-}term\ responsibility}}.
\]

We also use a conceptual measure of Trustworthy Change Productivity:
\[
\mathrm{TCP}=\frac{\text{Value of Accepted TC}}{\text{Complete engineering resources}}.
\]
Value may be represented by user outcomes, revenue, cost reduction, reliability, risk reduction, or public value. TCP is an organizational-economic construct, not a universal scalar suitable for comparing all software systems.

\paragraph{Primary metrics for Agent SE.}
Table~\ref{tab:metrics} summarizes the primary candidate measures used by the framework.
\begin{table}[H]
\centering
\small
\caption{Primary candidate metrics for Agent SE.}
\label{tab:metrics}
\begin{tabularx}{\linewidth}{@{}p{0.30\linewidth}X@{}}
\toprule
\textbf{Category} & \textbf{Example measures} \\
\midrule
Output & Accepted TC, TC Value \\
Digital resources & Token/TC, Compute/TC, Tool Cost \\
Human resources & Human Attention/TC, Review Fragmentation \\
Waste & Rejected Change Cost, Retry Token Ratio \\
Flow & Verification Queue, Acceptance Latency \\
Architecture & Artifact Span, Context Span, Responsibility Span \\
Team context & Context Staleness, Invalidation Precision/Recall, Stale-Context Escape Rate \\
Verification & Evidence Independence, Escaped Defect \\
Continuity & Takeover Time, Anchor-Absence Recovery Time \\
\bottomrule
\end{tabularx}
\end{table}

If the order-status split is measured only by coding time, it may show dramatic agent acceleration. The complete TC still requires clarification of late callbacks, definition of historical semantics, verification of payment invariants, legacy-client testing, analytics changes, anchor acceptance, and observation of production behavior. Coding acceleration describes Construction; Agent SE must observe the complete Trustworthy Change Flow.

\section{Research and Industrial Landscape as of September 2, 2026}
\label{sec:landscape}

\subsection{Evidence Mapping Scope}
The observation date for this section is frozen at
\[
\text{2 September 2026}.
\]
We distinguish evidence sources by the claims they can support. Table~\ref{tab:evidence-types} summarizes the role assigned to each evidence type in the landscape mapping.

\begin{table}[H]
\centering
\small
\caption{Evidence types and their roles in the landscape mapping.}
\label{tab:evidence-types}
\begin{tabularx}{0.9\linewidth}{@{}p{0.31\linewidth}X@{}}
\toprule
\textbf{Evidence type} & \textbf{Primary use} \\
\midrule
Peer-reviewed paper & Methods, empirical results, and research questions \\
Preprint / vision paper & Conceptual frameworks and propositions awaiting validation \\
Industrial paper & Reported deployment and observed outcomes \\
Official documentation & Whether a product capability exists at the freeze date \\
Vendor engineering practice & Engineering-method and infrastructure trends \\
Survey & Adoption, self-reported behavior, and organizational perception \\
\bottomrule
\end{tabularx}
\end{table}

The existence of a product capability establishes that the capability has entered practice. It does not establish a causal effect on software quality or productivity. Likewise, an arXiv framework may be close to this paper conceptually without having established empirical validity. We therefore use the landscape to locate constructs and open questions, not to turn proximity into evidence for our claims.

\subsection{Agentic SE, Responsibility Protocols, Specification, and Harnesses}
SE~3.0 and SASE advance intent-centric and agent-native views of software engineering and expand the set of actors, processes, tools, and artifacts under consideration~\cite{ref8,ref9}. MAGE makes representation, evidence, and authority explicit as engineering concerns when implementation becomes abundant~\cite{ref10}. A related case study grounds governance conversion in recurring failures observed during high-velocity agentic implementation and the durable controls created in response~\cite{ref59}. Feldt and colleagues' Semi-Executable Stack similarly widens the scope of software engineering beyond executable code toward instructional and operating artifacts whose interpretation still matters~\cite{ref36}. Hoda's ``beyond code'' framing argues for a whole-of-process view of Agentic SE rather than a coding-only one~\cite{ref44}. Alenezi characterizes the shift toward supervised agent workflows, verification bottlenecks, and outcome ownership~\cite{ref42,ref43}. These works are important prior context for our thesis; the novelty of this paper does not lie in claiming that Agent SE should extend beyond code, verification, or human oversight.

A newer line is closer to the responsibility boundary itself. Prifti's AI-SDLC protocol language formalizes human--agent responsibility boundaries, approval gates, capability boundaries, and process invariants~\cite{ref41}. GAIE calibrates human oversight intensity to regulatory impact, reversibility, customer proximity, and data sensitivity~\cite{ref49}. Risk Architecture for AI-Native Engineering Teams focuses on roles, decision rights, escalation, and assurance at the engineering-management layer~\cite{ref50}. These are adjacent to Responsibility Topology but not equivalent. They specify \emph{how responsibility is constrained or oversight is calibrated}; our classification asks \emph{how many independent residual-risk acceptance centers a particular software baseline or change actually requires under current governance}.

\paragraph{Specification and harness engineering.}
Spec-Driven Development became a visible practical trend in 2026. Microsoft argues for structured specifications as a shared engineering basis for humans and AI~\cite{ref21}; GitHub Spec Kit makes Spec, Plan, Task, Implementation, clarification, and analysis explicit workflow artifacts~\cite{ref22}. The Specification Paradox argues that code automation shifts complexity toward domain understanding, requirements, validation, and evolution~\cite{ref13}. Our Minimum Sufficient Specification hypothesis is deliberately narrower: it asks whether, for classes of changes, there is an empirically identifiable region beyond which added specification rigor ceases to repay its lifecycle cost.

Harnesses are also becoming an object of study rather than merely a product feature. A source-code study of eleven production coding-agent harnesses identifies recurring subsystems for loops, tools, context management, safety controls, orchestration, and extension surfaces~\cite{ref46}. This evidence supports the stance in Section~\ref{sec:tooling}: the methods layer of this paper is intentionally synthetic. Sandboxes, hooks, tools, context mechanisms, and policy guards are not claimed as new primitives; the contribution is to connect them to TC state, responsibility, and authoritative engineering semantics.

\subsection{Verification, Human--Agent Organization, and Economics}
Programming with Trust and Trustworthy AI Software Engineers make assurance, transparency, responsibility, and epistemic discipline central concerns~\cite{ref11,ref12}. AgentLens shows that final test success alone can conceal process failures and ``lucky passes''~\cite{ref14}. A September 2026 benchmark study further shows that nominal labels such as ``bug fix'' or ``feature'' are poor proxies for the actual demands placed on coding agents; its Spread--Novelty--Centrality profile separates benchmarks that share the same category label and links task demands to agent behavior~\cite{ref47}. This result is directly relevant to our research agenda: TC-oriented evaluation should characterize change scope, evidence, context, and acceptance demands rather than treating issue labels as sufficient task descriptors.

OpenAI's 2026 evaluation reports likewise discuss contamination, task quality, and signal/noise problems in coding benchmarks~\cite{ref26,ref27}. NIST treats confabulation as a governance risk~\cite{ref31}. At the execution-security boundary, Task Shield constrains instructions and tool calls through task alignment~\cite{ref39}, while MCPTox demonstrates systematic tool-poisoning risk in real MCP servers~\cite{ref40}. These lines motivate a shift from Final Pass toward claims, evidence, process quality, execution security, and evaluation validity.

\paragraph{Human--Agent organization and observable responsibility boundaries.}
AIDev observes coding agents in real GitHub development~\cite{ref15}; normative work on Human--AI Software Engineering Teams uses concepts such as commitment, obligation, prohibition, and permission~\cite{ref19}. OrgAgent, company-like heterogeneous-agent organizations, and self-organizing development systems organize the agent workforce in different ways~\cite{ref16,ref17,ref18}. A mixed-methods study of 448 professional developers at Microsoft found that accepted AI autonomy varies by task and individual, and that task accountability is associated with lower odds of allowing AI to act on a developer's behalf~\cite{ref45}. That finding does not validate Responsibility Topology, but it demonstrates that accountability and autonomy boundaries are observable properties of real software work rather than purely normative abstractions.

Cross-organizational deployment introduces another boundary. Reid and colleagues distinguish singular, federated, and open governance for multi-agent interactions across organizational perimeters~\cite{ref48}. Their framework is broader than software-change acceptance, but it provides a useful external coordinate for the federated extensions discussed in Section~\ref{sec:discussion}.

\paragraph{Project management and bounded capacity.}
Agentic Software Project Management has entered the ICSE research agenda~\cite{ref20}, while METR's evolving productivity methodology emphasizes that concurrency, task selection, and agent use complicate simple wall-clock comparisons~\cite{ref28}. These developments fit the bounded-capacity view: execution can expand rapidly while verification, attention, integration, and budgets remain finite.

\subsection{Industrial Practice and Comparative Positioning}
Industrial practice increasingly makes responsibility and verification explicit rather than treating agents only as coding tools. EPAM's Agentic Development Lifecycle includes human--agent responsibility mapping, approval gates, authority levels, and continuous governance as first-class lifecycle activities~\cite{ref57}. McKinsey's August 2026 synthesis of agentic product-development practice similarly highlights operating-model redesign, roles and responsibilities, verification mechanisms, AI operations, and organizational change as recurring characteristics of teams reporting greater impact~\cite{ref58}. These sources are practice evidence, not proof of our topology, but they reinforce the engineering relevance of responsibility boundaries.

Across OpenAI Codex, Anthropic Claude Code, GitHub Copilot/Spec Kit, Replit, Huawei CodeArts Agent, and related platforms, persistent instructions, specification artifacts, skills, hooks, MCP tools, sandboxes, subagents, parallel workspaces, context layers, governance controls, and cost controls are increasingly routine~\cite{ref23,ref24,ref25,ref29,ref30}. A broad evolution can be summarized as
\[
\text{Prompt}\rightarrow\text{Harness}\rightarrow\text{Agent Runtime}\rightarrow\text{Team/Organization Control}.
\]
The point is not that every product implements the theoretical constructs in this paper. Rather, the implementation substrate for testing those constructs already exists. Table~\ref{tab:adjacent-governance} summarizes the closest adjacent 2026 governance directions.

\begin{table}[H]
\centering
\small
\caption{Adjacent 2026 research on responsibility and governance. ``Adjacent'' means conceptually close, not equivalent to Responsibility Topology.}
\label{tab:adjacent-governance}
\begin{tabularx}{\linewidth}{@{}p{0.23\linewidth}p{0.34\linewidth}X@{}}
\toprule
\textbf{Direction} & \textbf{Primary object} & \textbf{Relation to this paper} \\
\midrule
AI-SDLC protocols~\cite{ref41} & Human--agent boundaries, approval gates, process invariants & Formalizes responsibility protocols; does not classify organizations by independent residual-risk acceptance centers \\
Graduated oversight~\cite{ref49} & Oversight intensity conditioned on risk and regulation & Calibrates how much oversight a task receives; orthogonal to how acceptance authority is distributed \\
Risk architecture~\cite{ref50} & Roles, decision rights, escalation, framework adequacy & Closely related governance layer; broader than TC-specific acceptance topology \\
Whole-process Agentic SE~\cite{ref44,ref42,ref43} & Lifecycle scope, workflow supervision, accountable judgment & Establishes prior whole-process and accountability framing that this paper builds upon \\
Harness engineering~\cite{ref46} & Runtime architecture of production coding agents & Provides empirical substrate for our tooling projection; not itself a responsibility model \\
\bottomrule
\end{tabularx}
\end{table}

Table~\ref{tab:coverage} provides a broader qualitative construct-coverage comparison used only for theoretical positioning.

\begin{landscape}
\begingroup
\scriptsize
\setlength{\tabcolsep}{2pt}
\renewcommand{\arraystretch}{1.22}
\begin{longtable}{@{}L{0.115\linewidth}L{0.125\linewidth}L{0.120\linewidth}L{0.125\linewidth}L{0.160\linewidth}L{0.145\linewidth}L{0.170\linewidth}@{}}
\caption{Qualitative construct-coverage comparison of major research directions. Entries are author judgments for theoretical positioning, not results of a coded systematic review or quantitative mapping.}\label{tab:coverage}\\
\toprule
\textbf{Dimension} & \textbf{SE 3.0 / SASE} & \textbf{MAGE} & \textbf{SDD / Harness} & \textbf{Trust / governance} & \textbf{Agent organization} & \textbf{This work} \\
\midrule
\endfirsthead
\toprule
\textbf{Dimension} & \textbf{SE 3.0 / SASE} & \textbf{MAGE} & \textbf{SDD / Harness} & \textbf{Trust / governance} & \textbf{Agent organization} & \textbf{This work} \\
\midrule
\endhead
\bottomrule
\endfoot
Agent as first-class actor & Explicitly modeled & Explicitly modeled & Explicitly modeled & Explicitly discussed & Principal organizational unit & \textbf{Inherited and integrated} \\
Intent / Spec & First-class concern & Purposeful representation & First-class artifact & Supports assurance & Partially addressed & \textbf{Progressive Specification} \\
Harness / Permission & Addressed & Addressed & First-class mechanism & Approval/capability boundaries addressed & Addressed & \textbf{Linked to TC and responsibility} \\
Evidence & Structured artifacts such as MRP & Validators / gates & Quality gates & First-class assurance concern & Reviewer roles & \textbf{First-class TC component} \\
Responsibility Topology & Not a primary classification axis & Human authority represented; topology not explicitly classified & Not a first-class object & Accountability, protocols, oversight, decision rights & Governance mainly organizes agent work & \textbf{Independent residual-risk acceptance as primary axis} \\
HAC & Human--AI collaboration modeled more broadly & Not explicitly modeled & Not explicitly modeled & Human/AI roles discussed & Agent is principal organizational unit & \textbf{First-class execution unit} \\
Unified TC object & Partial conceptual overlap & Per-change obligations & Spec-to-code workflow & Assurance attached to processes/actions & Not change-centered & \textbf{First-class engineering object} \\
Shared Authoritative State & Not a team-level first-class object & Partial conceptual overlap & Spec as source of truth & Governance state partially addressed & Shared memory/state addressed & \textbf{Explicit authority, versioning, and invalidation} \\
Context Invalidation & Not explicit as a team protocol & Not a primary mechanism & Specification evolution supported & Not primary & Memory/communication addressed & \textbf{Explicit protocol} \\
Delegation / Responsibility graphs & Not explicitly separated & Partial overlap & Workflow-centric & Responsibility boundaries and escalation addressed & Delegation structure central & \textbf{Explicitly separated} \\
Accepted-TC Economics & Not a primary accounting object & Cost as constraint & Efficiency/cost discussed & Not primary & Token/cost experiments & \textbf{Explicit research agenda} \\
\end{longtable}
\endgroup
\end{landscape}

As of the freeze date, the most important open questions for this framework remain empirical: whether Responsibility Topology adds explanatory power beyond team size, whether Minimum Sufficient Specification can be identified in practice, whether Context Invalidation reduces stale-context failures, and whether Accepted-TC-centered accounting improves measurement of agentic software work.

\section{Empirical Commitments and a Falsifiable Research Agenda}
\label{sec:agenda}

\subsection{Operationalization and Validation Contract}
The constructs introduced in this paper have sustained value as software-engineering theory only if they produce empirically distinguishable expectations.

This section further operationalizes Responsibility Topology, Human--Agent Cell, Trustworthy Change, Context Coherence, Progressive Specification, and Bounded-Capacity Agent SE by connecting abstract constructs to observable variables, directional hypotheses, and falsification conditions.

The paper covers the two precursor stages of theory development: \textbf{conceptual development}, which defines constructs and relations, and \textbf{operationalization}, which maps those constructs onto observable engineering phenomena. Controlled experiments, field studies, and longitudinal studies remain responsible for empirical testing. The operationalization follows a goal--question--metric logic: the theoretical claim is stated first, then research questions and observables are selected so that support and failure can be distinguished~\cite{ref37}.

This section therefore states the framework's \textbf{empirical commitments}. Each major claim should answer three questions: how can the construct be observed or measured; what relation or difference should appear if the claim holds; and what result should weaken, modify, or reject the claim?

Recent empirical and benchmark work provides useful signals without validating the framework in advance. In a study of 448 professional developers, task accountability was associated with lower willingness to let AI act on the developer's behalf~\cite{ref45}; this motivates, but does not prove, the proposition that responsibility structure affects autonomy boundaries. Likewise, recent benchmark analysis shows that nominal categories such as ``bug fix'' or ``feature'' can hide materially different task demands~\cite{ref47}. These findings argue for measuring the constructs proposed here directly rather than treating existing task labels or headcount as sufficient proxies.

\paragraph{Core empirical commitments.}
Table~\ref{tab:empirical-commitments} states the core propositions, observable expectations, and weakening signals.
\begin{table}[H]
\centering
\scriptsize
\caption{Core theoretical claims and falsification signals.}
\label{tab:empirical-commitments}
\begin{tabularx}{\linewidth}{@{}p{0.06\linewidth}p{0.28\linewidth}p{0.31\linewidth}X@{}}
\toprule
\textbf{ID} & \textbf{Theoretical claim} & \textbf{Observable expectation} & \textbf{Main weakening or falsification signal} \\
\midrule
P1 & Minimum Sufficient Specification occupies a task-dependent effective region & Raising rigor from an under-specified level reduces semantic divergence, rework, or escaped specification defects; later marginal benefit slows while specification cost continues to rise & After controlling task type and model/harness, additional rigor does not systematically reduce rework, semantic divergence, or escaped specification defects over any practically relevant range; or incremental specification cost is consistently offset by downstream savings, leaving no identifiable minimum-sufficient region \\
P2 & Agent concurrency is constrained by verification capacity & With fixed verification resources, Accepted-TC throughput eventually saturates or falls & Throughput remains near-linear while verification burden does not increase \\
P3 & Evidence Independence improves assurance & Greater independence is associated with lower escaped-defect rates & Changes in independence have no stable relation to escaped defects \\
P4 & Change Locality affects complete engineering cost & Larger Artifact, Context, or Responsibility Span predicts higher cost or latency & The spans add no explanatory power for cost or quality \\
P5 & Context Invalidation improves team coherence & Appropriate invalidation reduces stale-context escapes & Invalidation only increases interruption without reducing semantic errors or rework \\
P6 & Responsibility Topology has independent explanatory power & After controlling headcount and system scale, topology still predicts differences in acceptance latency, assurance, or continuity & Topology adds no explanatory power after covariate control \\
P7 & Delegation--Responsibility Misalignment creates governance risk & More misalignment predicts unauthorized change, rollback, or responsibility ambiguity & Misalignment measures have no stable relation to these outcomes \\
P8 & Takeoverability reflects responsibility continuity & Higher Takeoverability predicts lower transfer time and recovery cost & Takeoverability does not explain continuity outcomes \\
\bottomrule
\end{tabularx}
\end{table}

These propositions are deliberately exposed to evidence that may force the framework to narrow or change.

\subsection{Specification, Ambiguity, Effort, and Capacity}
\paragraph{Minimum Sufficient Specification.}
Multiple specification levels can be constructed for the same change. For example, S0 may be a one-sentence natural-language request; S1 adds normal examples; S2 adds counterexamples and non-goals; S3 uses a state machine or decision table; S4 uses executable contracts or formal properties.

Candidate Divergence, Clarification Count, Token use, Verification Effort, Rework, and Escaped Defect can then be measured to study whether
\[
C_{\mathrm{spec}}=C_{\mathrm{specification}}+C_{\mathrm{generation}}+C_{\mathrm{verification}}+C_{\mathrm{rework}}+C_{\mathrm{invalidation}}
\]
exhibits a task-dependent effective region.

\paragraph{Ambiguity proxy metrics.}
Ambiguity is difficult to observe directly. Candidate proxies include Clarification Question Count, Unresolved Specification Slots, Mutually Exclusive Interpretations, Candidate Semantic Divergence, and Replan Rate induced by semantic clarification.

Clarification Count alone is insufficient because models and harnesses differ in how readily they ask questions. Empirical work should calibrate proxies within model/harness conditions and compare their explanatory power for rework, candidate divergence, and verification cost.

\paragraph{Value of clarification.}
Strategies such as Always Ask, Never Ask, Confidence-Based Ask, and Risk-Based Ask can be compared. Define
\[
\mathrm{VoC}=\text{Expected loss without clarification}-\text{Cost of clarification}.
\]
The main question is which combinations of ambiguity and risk justify interrupting agent execution to obtain a human semantic decision.

\paragraph{Concurrency, effort, and topology economics.}
\paragraph{Concurrency turning point.}
Increase the number of parallel agents while controlling Task, Model, and Harness, and measure Candidate Rate, Accepted-TC Rate, Verification Queue, Human Attention, Rework, and Token/Accepted TC. If trustworthy throughput first rises and then saturates or declines, further work can estimate an effective concurrency region conditioned on Task Decomposability.

\paragraph{Agentic effort estimation.}
Study whether Specification Ambiguity, Context, Change Locality, Tool Steps, Risk, Verification Requirement, and Responsibility Topology jointly explain or predict Token use, Human Attention, Lead Time, and Rework. Statistical and machine-learning estimators can then be compared for cross-project generalization. No particular functional form is assumed in advance.

\paragraph{Topology economic pressure.}
Study whether the $C_{\mathrm{single}}$ and $C_{\mathrm{multi}}$ terms from Section~\ref{sec:metrics} can be reliably estimated and how Verification Queue, Anchor Attention, and Context Synchronization Cost influence organizational choices. Economic pressure must be separated from Hard Governance Triggers so that a Responsibility Anchor is not reduced to a throughput resource.

\subsection{Evidence, Architecture, and Context}
Compare Generator Self-Check, Same-Model Reviewer, Different-Model Reviewer, Deterministic Tool, Cross-HAC Verification, and Domain Expert verification. Outcomes include defect detection, specification omission, and escaped defects.

Evidence independence can be decomposed into Model, Context, Criterion, and Responsibility Independence. The research question is which dimensions actually reduce common-mode failure. If different-model review does not reduce correlated errors after controlling other factors, model diversity alone should not be treated as a valid independence proxy.

\paragraph{Architecture and context coherence.}
\paragraph{Change locality.}
Collect Artifact Span, Context Span, and Responsibility Span separately and analyze their relationships with Token use, Review, Revalidation, Lead Time, and Defect outcomes.

\paragraph{Consistency policy.}
Compare Strong Consistency, Eventual Consistency, and Local State for different engineering facts. Evaluate how State Type, Risk, Dependency, and Change Frequency explain which policy is appropriate. The framework does not assume that these factors collapse into a known scalar function.

\paragraph{Context invalidation.}
Use expert judgment as a reference to evaluate automatic selection among Continue, MarkStale, Refresh, Replan, Reverify, and Stop. Measures include Invalidation Precision, Invalidation Recall, Unnecessary Interruption, Missed Invalidation, Stale-Context Escape Rate, and Downstream Defect. These measures distinguish Context Coherence evaluation from benchmarks that observe only final pass/fail.

\subsection{Responsibility Topology and Continuity}
\paragraph{Explanatory power of responsibility topology.}
Compare single-anchor and multi-anchor systems while controlling Domain, System Size, Team Size, and Risk. Outcomes can include Decision Latency, Accepted-TC Throughput, Escaped Defect, Incident Response, Responsibility Ambiguity, Takeover Time, and Anchor-Absence Recovery Time.

\paragraph{Delegation--responsibility misalignment.}
Define misalignment patterns such as an executor without a responsible anchor, a required anchor missing from acceptance, permissions extending beyond the responsibility domain, or a release gate omitting necessary acceptance. Study whether these patterns predict Unauthorized Change, Rollback, or Incident outcomes.

\paragraph{Augmented Conway effect.}
Collect the Human Communication Graph, Agent Delegation Graph, Context Dependency Graph, Module Dependency Graph, and Co-change Graph. Project the graphs onto a shared Task, Module, or TC space, then test whether Agent Delegation and Context Dependency add explanatory power after controlling Human Communication.

Near-term engineering outcomes should be preferred: Rework, Integration Conflict, Context Revalidation, TC Lead Time, and Defects. The hypothesis does not presuppose that graph misalignment directly causes any specific API failure.

\paragraph{Human attention and continuity.}
\paragraph{Attention fragmentation.}
Record Human Hours together with Context Switches, Deep-Work Duration, Review Fragmentation, and Approval Count. Study the point at which more agent concurrency begins to increase human cognitive fragmentation.

\paragraph{Blind takeover.}
Ask a qualified subject who did not participate in daily development to Identify the Baseline, Stop Agents, Deploy a Fix, Restore Service, Explain the Architecture, and Assume Responsibility. Measures include Takeover Success, Takeover Time, Undocumented Dependencies, and Privilege Recovery Time.

\paragraph{Anchor-absence resilience.}
Simulate temporary unavailability of the primary anchor without formal responsibility transfer. Observe whether the system can degrade automatically, block high-risk changes, continue pre-authorized low-risk tasks, and wait safely for the anchor to return.

\subsection{Benchmark and Reporting Protocol}
Future benchmarks can expand from
\[
\text{Issue}\rightarrow\text{Patch}\rightarrow\text{Test Pass}
\]
to
\[
\text{Intent}\rightarrow\text{Clarification}\rightarrow\text{Candidate}\rightarrow\text{Evidence}\rightarrow\text{Acceptance}\rightarrow\text{Runtime}.
\]
Evaluation would then cover Correctness, Scope, Permission, Evidence, Context Handling, Human Attention, Cost, and Responsibility Closure.

A Single-Center Track should emphasize single-anchor bandwidth, external assurance, Takeoverability, and Anchor-Absence Resilience. A Multi-Anchor Track should emphasize effective acceptance sets, joint acceptance, and Responsibility Closure; studies with multiple HACs under either topology should additionally report context coherence, cross-cell verification, and Context Invalidation.

Agent SE studies should report Model, Version, Harness, Tools, Permissions, Context, Budget, and Experiment Date. Benchmark descriptions should also report the demand profile of the change itself rather than relying only on labels such as ``bug fix'' or ``feature''~\cite{ref47}. Table~\ref{tab:evaluation-layers} summarizes the resulting evaluation layers and primary records.

\begin{table}[H]
\centering
\small
\caption{Evaluation layers for Agent SE research.}
\label{tab:evaluation-layers}
\begin{tabularx}{\linewidth}{@{}p{0.26\linewidth}X@{}}
\toprule
\textbf{Layer} & \textbf{Primary records} \\
\midrule
Task & Type, Ambiguity Proxy, Risk, Context, Permission \\
Execution & Model, Token, Tool, Retry, Human Intervention \\
Change & Candidate, R1--R6, Evidence, Eligible, Accepted \\
Team & Context Version, Anchor, Cross-cell Verification, Invalidation Precision/Recall, Stale-Context Escape \\
Lifecycle & Runtime, Defect, Maintenance, Long-Term Responsibility Obligations \\
Continuity & Takeover, Anchor Transition, Anchor-Absence Recovery \\
\bottomrule
\end{tabularx}
\end{table}

The purpose of these designs is to expose the core constructs of the framework to empirical risk rather than leaving them as unfalsifiable conceptual labels.

\section{Discussion and Conclusion: What Software Engineering Governs in the Agent Era}
\label{sec:discussion}

\subsection{Organizational and Economic Implications}
Agents substantially increase the amount of software execution that one responsibility center can mobilize. This shift may allow very small human cores to develop and operate more products for longer periods.

Multi-person responsibility structures remain necessary when software involves independent professional responsibility, separation of duties, legal or regulatory requirements, or continuous operational obligations.

Earlier service-ecosystem research provides a related pre-agent precedent for coordinating multiple stakeholders and independently controlled providers under shared objectives~\cite{ref53,ref54}. Agents change the feasible range of different Responsibility Topologies. Single-center and multi-anchor forms are two basic responsibility structures jointly shaped by software risk, agent capability, organizational conditions, and institutional environment, not two stages of organizational size.

\paragraph{The changing value of teams.}
One traditional value of software teams is additional labor. Once one HAC can already invoke multiple agents in parallel, the marginal value of a new HAC increasingly lies in independent knowledge, independent context, independent evidence, and independent acceptance.

The engineering value of a team depends more strongly on genuine heterogeneity. If multiple HACs use the same model, the same context, and the same implicit assumptions, headcount does not automatically increase assurance.

An important design goal for future Team SE is to preserve professional difference, independent rejection rights, and responsibility continuity under high digital execution capacity.

\paragraph{Risk-adaptive autonomy.}
Agents can perform longer tasks, call more tools, and act in production. These capabilities expand execution radius, but software-engineering maturity should not be measured by autonomy alone.

More appropriate dimensions include task scope, context validity, permission, budget, evidence, stop mechanisms, recovery, and responsibility. Autonomy should be adapted to observable properties such as impact, reversibility, detectability, available evidence, and responsibility structure. No fixed function over these factors is assumed; autonomy should be jointly constrained by risk-relevant properties.

At the opposite extreme, requiring humans to approve every important action can itself create verification bottlenecks and rubber-stamping. Human attention should be concentrated where semantic commitment and residual-risk acceptance are actually required.

\subsection{Engineering and Educational Implications}
As natural language becomes a major development interface, prompting can appear to become the central engineering skill. Yet once interaction contains state machines, interfaces, invariants, exceptions, and acceptance criteria, it has entered the domain of specification.

Future engineers may write fewer program statements directly, but they still need to understand program behavior, state, architecture, security, performance, failure, and data as causal structures. Agent SE education should preserve enough computing and software foundations to support independent judgment while adding agent-native abilities in intent, specification, delegation, evidence, integration, operation, and responsibility.

Professional value will increasingly lie in deciding what should change, constraining digital execution, identifying the evidence on which change may be accepted, and knowing when to reject or stop.

\paragraph{Architecture and organizational memory.}
The ability of agents to process larger contexts does not make modularity obsolete. Cross-module change still raises context, token, verification, and responsibility-closure costs. A central objective of agent-oriented architecture is:
\begin{quote}
Enable most changes to close within bounded context and bounded responsibility scope.
\end{quote}

Organizational memory faces a similar transition. Future software organizations will accumulate large volumes of chat, traces, documents, decisions, commits, and incident records. As retrieval improves, the harder question shifts from ``can we find the information?'' to ``is the information we found still authoritative?'' Thus
\[
\text{Knowledge Retrieval}\rightarrow\text{Authority Management}
\]
may become a foundational infrastructure shift in agentic software organizations.

\subsection{Scope, Generalization, and Limitations}
Agents can progressively assume coding, testing, debugging, migration, deployment, monitoring, and diagnosis. High-impact action still raises three questions: who allowed the action; who judges the result reliable enough; and who bears the residual responsibility when consequences occur?

Many responsibilities can be encoded in advance through policy, pre-authorization, contracts, and deterministic controls. Yet creating those policies, setting risk thresholds, and changing the rules still require explicit Responsibility Anchors. Responsibility Topology provides an organizational axis independent of human headcount and agent count.

\paragraph{Generalizing the topology.}
This paper uses single-center and multi-independent-center forms as two basic types. The empirical commitments in Section~\ref{sec:agenda} cover only these types and their directly derived mechanisms; they do not attempt to validate more complex topologies.

Real organizations may include hierarchical responsibility, federated responsibility, temporary joint acceptance, external certified authority, or platform-mediated responsibility. Work on multi-agent deployment across organizational boundaries already distinguishes singular, federated, and open governance regimes~\cite{ref48}. Our two-type classification does not subsume those regimes. It provides a local building block that asks, for a particular TC inside a given governance regime, which independent anchors must accept the residual risk. Enterprise and supply-chain extensions remain a separate research problem outside the present empirical claims.

\paragraph{What is software engineering ultimately engineering?}
As more programs are generated by agents, the center of Software Engineering shifts from code production toward \textbf{trustworthy software change}.

Intent determines why change is needed. Specification constrains its allowable semantics. Delegation determines who can act and under what resource boundary. Evidence justifies claims. Integration turns local changes into a shared system. Operation continues to test the assumptions on which acceptance depended. Responsibility ensures that software consequences remain traceable to subjects with legitimate authority to decide and remediate.

From this perspective, Agent-era Software Engineering can be defined as:
\begin{quote}
\textbf{The organization of humans and scalable digital execution under bounded resources and explicit responsibility so that software changes can be continuously formed, verified, integrated, accepted, operated, and sustained through their lifecycles.}
\end{quote}

\subsection{Conclusion}
Agentic execution changes the balance of software engineering more than it changes its purpose. Code, tests, migrations, and operational analyses can now be produced at a rate that was implausible when most software-process theory was formed. The difficult question is what happens after that execution has been produced: which intent it implements, which evidence justifies it, which other work it invalidates, and who is authorized to accept the residual risk.

This paper has organized that problem around two primary constructs and one execution abstraction. A Human--Agent Cell is the execution boundary. Trustworthy Change is the change-centered engineering object. Responsibility Topology describes the distribution of independent residual-risk acceptance authority. None of these claims makes change management, accountability, specification, verification, or human oversight new. The intended contribution is a narrower reorganization of those concerns for a setting in which execution is cheap enough to stop being a reliable proxy for engineering completion.

The distinction between single-center and multi-anchor organizations matters because it changes formal closure conditions. Distributed HAC execution may already require authoritative shared state and invalidation when binding facts change. Multi-anchor governance makes the authority/version semantics of that state, cross-domain evidence, and explicit Responsibility Closure across independent anchors structurally necessary when closure spans responsibility domains.

This framework is not a finished empirical theory. Its value now depends on whether its distinctions survive measurement. Responsibility Topology may add no explanatory power after team size and risk are controlled. A minimum-sufficient specification region may not appear. Evidence diversity may fail to reduce common-mode error, and context invalidation may cost more than it saves. Any of those results would narrow or remove part of the framework; that is the point of stating the commitments explicitly.

A compact summary is
\[
\boxed{HAC\rightarrow\text{Candidate Change}\rightarrow\text{Eligible Change}\rightarrow\text{Accepted Trustworthy Change}}.
\]
The agent era increases the capacity to produce the left side of this chain. Software engineering remains responsible for whether the right side is reached for reasons that can be inspected and challenged, with the resulting obligations sustained by an identifiable organization.

\clearpage
\printbibliography[title={References}]

\end{document}